\documentclass[reprint,aps,pra,groupedaddress,superscriptaddress]{revtex4-1}

\usepackage{graphicx}
\usepackage{bm}
\usepackage{xcolor}
\usepackage{amsmath}
\usepackage{braket}
\usepackage[flushleft]{threeparttable}
\usepackage{xcolor}
\usepackage{color}
\usepackage{colortbl}
\usepackage{multirow}
\usepackage{hhline}
\usepackage{soul}
\let\vec\mathbf
\definecolor{tcolor}{RGB}{250,200,200}  
\definecolor{zcolor}{RGB}{200,200,250}  
\begin{document}
\title{The missing link between standing- and traveling-wave resonators}  

\author{Qi Zhong}
\email[]{qizhong@mtu.edu}
\affiliation{Department of Physics, Michigan Technological University, Houghton, Michigan 49931, USA}

\author{Haoqi Zhao}
\affiliation{Department of Electrical and Systems Engineering, University of Pennsylvania, Philadelphia, Pennsylvania 19104, USA}

\author{Liang Feng}
\affiliation{Department of Materials Science and Engineering, University of Pennsylvania, Philadelphia, Pennsylvania 19104, USA}

\author{Kurt Busch}
\affiliation{Humboldt-Universit\"{a}t zu Berlin, Institut f\"{u}r Physik, AG Theoretische Optik \& Photonik, D-12489 Berlin, Germany}
\affiliation{Max-Born-Institut, Max-Born-Stra{\ss}e 2A, 12489 Berlin, Germany}

\author{\c{S}ahin K. \"{O}zdemir}
\affiliation{Department of Engineering Science and Mechanics, The Pennsylvania State University, University Park, Pennsylvania 16802, USA}
\affiliation{Materials Research Institute, The Pennsylvania State University, University Park, Pennsylvania 16802, USA}

\author{Ramy El-Ganainy}
\email[]{ganainy@mtu.edu}
\affiliation{Department of Physics, Michigan Technological University, Houghton, Michigan 49931, USA}
\affiliation{Department of Electrical and Computer Engineering, Michigan Technological University, Houghton, Michigan 49931, USA}

\begin{abstract}
Optical resonators are structures that utilize wave interference and feedback to confine light in all three dimensions. Depending on the feedback mechanism, resonators can support either standing- or traveling-wave modes. Over the years, the distinction between these two different types of modes has become so prevalent that nowadays it is one of the main characteristics for classifying optical resonators. Here, we show that an intermediate link between these two rather different groups exists. In particular, we introduce a new class of photonic resonators that supports a hybrid optical mode, i.e. at one location along the resonator the electromagnetic fields associated with the mode feature a purely standing-wave pattern, while at a different location, the fields of the same mode represent a pure traveling wave. The proposed concept is general and can be implemented using chip-scale photonics as well as free-space optics. Moreover, it can be extended to other wave phenomena such as microwaves and acoustics.
\end{abstract}

\maketitle

\section{Introduction}
Light is a very peculiar form of energy that constantly travels from one point to another, which makes it difficult to store or `freeze' it in one place. This, however, can be effectively overcome by using optical resonators that utilize feedback mechanisms together with wave interference effects to recycle light along periodic trajectories. Depending on the resonator's geometry, these trajectories may intercept each other in opposite directions forming standing-wave patterns with a vanishing Poynting vector. Alternatively, they may form closed loops that support degenerate circulating traveling-wave modes (clockwise (CW) or counterclockwise (CCW)) with a non-vanishing Poynting vector along the loop direction. This ability to confine and trap light has enabled several scientific breakthroughs over the past few decades.  Importantly, recent technological progress in micro- and nano-fabrication has enabled the realization of on-chip optical resonators with spatial dimensions comparable to the wavelength of the trapped light, or even smaller \citep{Iwanaga-PR} with a wide range of applications including microlasers \cite{Loncar2002APL,Wu2004APL,McCall1992APL,Sandoghdar1996PRA,Miao2016S,Zhang2020LSA,Peng2016PNAS} and sensing \cite{Foreman2015AOP,Yalcin2006IEEE,Wiersig2016PRA,Chen2017N,Hodaei2017N,Zhong2019PRL},  just to mention few examples. Despite the large variety in their designs (microrings, microdisks, photonic crystals, Bragg structures, etc.), sizes, and material systems, optical resonators are typically classified into one of the aforementioned categories, i.e. standing- or traveling-wave devices \cite{Vahala-OM,Vahala2003N,Saleh-FP}. This classification is generally accepted as complete. Thus, research in the field of optical resonators has focused on applications of these resonators and implementations of novel designs with unique features. In particular, standing-wave resonators can be engineered to support small mode volumes and high quality factors, which makes them perfect choice for engineering quantum light-matter interactions \cite{John1990PRL,Chang2006PRL,Louyer2005PRA,Moreau2001APL,Vuckovic2003APL,Bennett2005APL,Strauf2007NPhot,Pelton2002PRL,Zhong2021PRR}. On the other hand, traveling-wave resonators are the preferred platform for (classical and quantum) nonlinear optics due to the ability to engineer the interaction between different wave components and the unidirectional propagation properties of these modes which facilitates the input and output coupling \cite{Kues2019NPhot,Kippenberg2018S}. In addition, it was shown previously that the interaction between an atom and a standing-wave pattern of light depends on whether the latter is generated in a standing-wave resonator or as a result of interference between counter-propagating waves in a traveling-wave resonator \cite{Shore1991JOSAB}.

In this work we show that this classification scheme for optical resonators (as standing- or traveling-wave resonators) is not complete. Instead, we reveal a new type of optical modes supported by certain resonator structures that represents a missing link between these two categories. Specifically, we propose a new resonator concept that supports an optical mode exhibiting hybrid standing- and traveling-wave patterns simultaneously.

\begin{figure*}[!t]
	\includegraphics[width=6.2in]{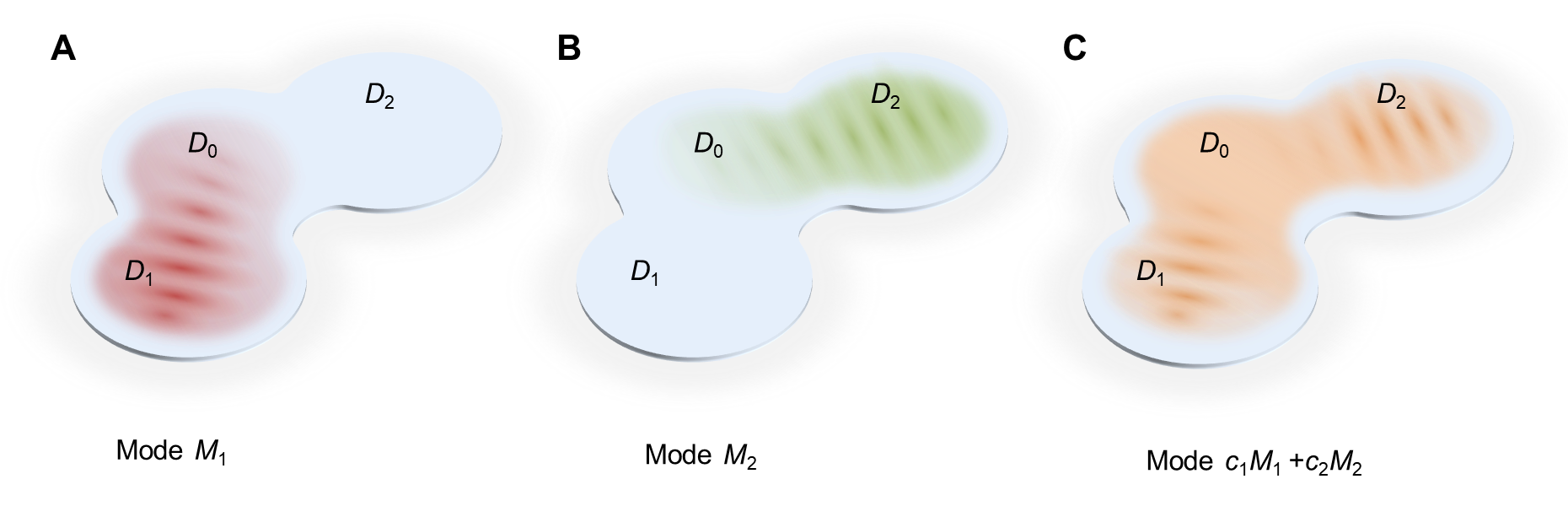}
	\caption{The concept of a hybrid-wave resonator. A resonator structure that supports two  degenerate standing-wave modes such that (A) mode $M_1$ resides in domain $D_0 \cup D_1$, and (B)  mode $M_2$ resides in domain $D_0 \cup D_2$. (C) It is possible that a proper linear superposition between $M_1$ and $M_2$ can result in new modes that preserve the standing-wave character in domains $D_{1,2}$ while at the same time form a traveling wave in domain $D_0$. In the figure, standing and traveling waves are schematically represented by interference fringes and uniform field distributions respectively.}
	\label{Fig-Schematic}
\end{figure*}

This article is organized as follows. First, we introduce a general concept outlining the behavior of the proposed resonator without a reference to a particular structure. Afterwards, we discuss an implementation based on standard chip-scale photonics technology. To gain insight into the modal structure of the proposed resonator, we present a detailed analysis of its modal features using a scattering matrix approach. Next, we confirm our results by using full-wave finite element simulations. Finally, we investigate the effect of local perturbation due to a small scatterer on the eigenmodes of the proposed hybrid-wave resonator.

\section{General concept} 
We start by considering a generic concept of optical resonators that supports modes overlapping only partially with the physical structure of the resonator. For simplicity, assume that such a resonator can be divided into three domains (extension to more domains is straightforward): $D_0$, $D_1$ and $D_2$ (the exact definition of the domain boundaries is not important). Furthermore, assume that it supports two degenerate standing-wave modes $M_{1,2}$ such that the field distribution of $M_1$ mainly resides in $D_0 \cup D_1$, and similarly the field associated with mode $M_2$ resides in $D_0 \cup D_2$ as shown schematically in Figure \ref{Fig-Schematic}A and \ref{Fig-Schematic}B respectively. The continuity conditions for the electromagnetic field across domain boundaries dictate that the field distributions associated with modes $M_{1,2}$ must be different across domain $D_0$. Let us now consider a general superposition of these two modes. It may be anticipated that, a new set of basis (owing to the degeneracy, there are infinitely many bases) can be constructed such that the standing-wave nature of modes $M_{1,2}$ in domains $D_{1,2}$ remain unaltered and yet the field distribution in $D_0$ forms a traveling-wave pattern due to a particular linear superposition of modes $M_{1,2}$ (Figure \ref{Fig-Schematic}C), akin to the relation $\cos(kz)+i \sin(kz)=\exp(ikz)$. Such a `mutant' resonator, if it exists, will support a `mutant' optical mode that, in some properly chosen basis, exhibits purely standing- and traveling-wave patterns at the same time---a feature that, to the best of our knowledge, has never been discussed before. We will refer to such a resonator as a hybrid-wave resonator. So far we have kept the discussion abstract. In what follows, we show that this abstract concept can be implemented in realistic optical resonator designs. In the main text, we focus on chip-scale devices but we note that it is straightforward to extend the discussion to implementations using free-space optics.  

\section{Integrated photonics implementation} 

To demonstrate that the concept discussed above can be realized by standard optical components, here we consider an implementation based on integrated photonics as shown in Figure \ref{Fig-Resonator}A (possible realizations based on free-space optics are discussed in Appendix A). The structure consists of three open ring sections connected by two beam splitters (labeled as $\text{BS}_{1,2}$). The outer rings here act as Sagnac loop reflectors \cite{Li2019LPR}. We note that even though variants of this geometry have been considered before for building various optical devices for different applications such as sensing, lasing, and information processing \cite{Simova2005JOSAB, Das2003OE, Parared2022APR,Shamy2022P,Moslehi1984IEEE}, the peculiar feature that we highlight in this work has escaped attention.   

\begin{figure*}[!t]
	\includegraphics[width=5.4in]{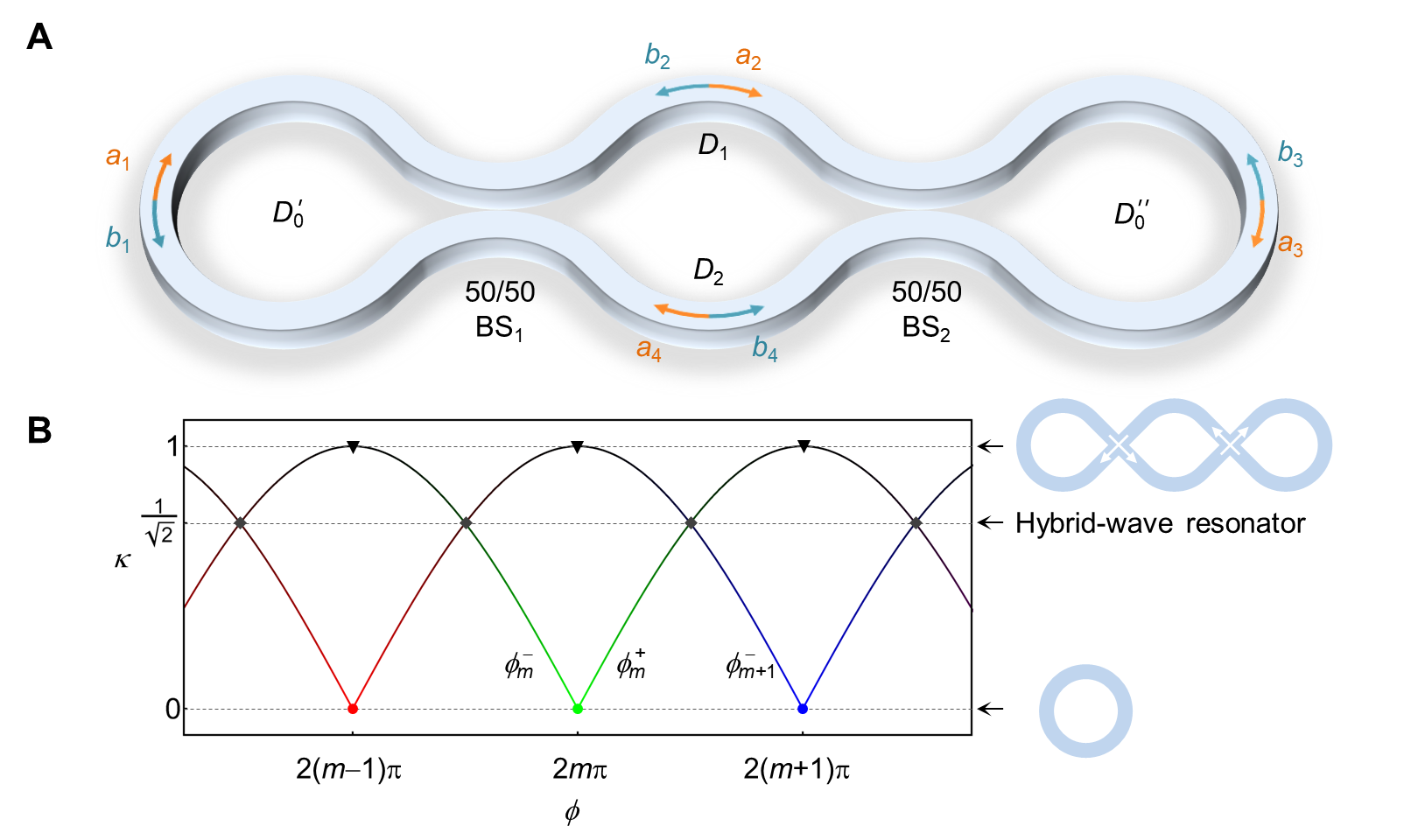}
	\caption{An implementation of a hybrid-wave resonator in integrated photonic platforms. (A) A hybrid-wave resonator can be constructed by deforming a ring resonator to introduce two 50/50 beam splitters, BS$_{1,2}$. The resonator can be divided into three domains: $D_0=D_0' \cup D_0''$ is the union of the left and right side rings, and  domains $D_1$ and $D_2$ represent the top and bottom middle sections, respectively. The field amplitudes at each location and their traveling directions are indicated on the figure and labeled as $a_i$ and $b_i$, $i=1,2,3,4$. (B) Resonant frequencies (horizontal axis) as a function of the beam splitter coupling coefficient $\kappa$ (vertical axis). Note that each $\kappa$ value represents an independent resonator structure. Horizontal dashed lines indicate values of $\kappa$ for which the spectrum is doubly degenerate.  The limit $\kappa=0$ corresponds to conventional microring resonator with two degenerate traveling-wave modes in clockwise and counterclockwise direction while the limit $\kappa=1$ corresponds to degenerate `knotted' modes. Hybrid-wave modes exist for $\kappa=1/\sqrt{2}$ as discussed in detail in the main text.}
	\label{Fig-Resonator}
\end{figure*}

To investigate the modal structure of this resonator, we will employ a scattering matrix analysis along the junctions indicated in Figure \ref{Fig-Resonator}A. Away from the beam splitter junctions, the field amplitudes can be decomposed into two traveling waves in opposite directions as shown in Figure \ref{Fig-Resonator}A. Within the context of scattering matrix formalism \cite{Yariv2000EL,Van-OMR,Saleh-FP}, the relations between these amplitudes are given by:
\begin{equation}\label{Eq-Matrix_equation}
	\begin{aligned}
\relax [a_1, b_1]^T = S_c [a_4, b_2]^T, \\
		[a_4, b_2]^T = S_c [a_3, b_3]^T, \\
		[a_3, b_3]^T = S_c [a_2, b_4]^T, \\
		[a_2, b_4]^T = S_c [a_1, b_1]^T,
	\end{aligned}
\end{equation}  
where $S_c=\exp{(i \phi/4)} S_b$ and $S_b =\begin{bmatrix}
	\tau & i \kappa\\
	i \kappa & \tau
\end{bmatrix}$ is the scattering matrix of each beam splitter. Here, $\tau$ and $\kappa$ are the field transmission and coupling coefficient of each beam splitter, and in the absence of any loss, they satisfy $\tau^2+\kappa^2=1$. For the special case of a 50/50 beam splitter, which is relevant to our discussion later, $\tau=\kappa=1/\sqrt{2}$. The phase term $\phi$ is defined as $\phi=2\pi n_\text{eff}Lf/c$, where $n_\text{eff}$ is the effective guiding index, $L$ is total length across the perimeter of the resonator, $f$ is the frequency and $c$ is the speed of light in vacuum. The numerical factor $1/4$ in the expression of $S_c$ arises because each wave component travels one quarter the length structure between any two consecutive junctions. In the absence of dispersion, $\phi$ is a linear function of $f$ and thus can be used directly to determine the resonant frequencies (the effect of dispersion is considered later in the full-wave simulations).

\begin{table*}[!t]
	\caption{Field components associated with the degenerate eigenmodes of the structure shown in Figure \ref{Fig-Resonator} as expressed in the three different bases $B_{1,2,3}$. } 
	\centering
	\begin{threeparttable}
	\begin{tabular}{|p{0.5in}<{\centering}|p{0.5in}<{\centering}|p{0.5in}<{\centering}|p{0.5in}<{\centering}|p{0.7in}<{\centering}|p{0.7in}<{\centering}|p{0.7in}<{\centering}|p{0.7in}<{\centering}|p{0.7in}<{\centering}|p{0.7in}<{\centering}|}
		\hline
	\multicolumn{2}{|p{1.0in}<{\centering}|}{Domains} & \multicolumn{2}{c|}{$D_0'$} & \multicolumn{2}{c|}{$D_1$} & \multicolumn{2}{c|}{$D_0''$} & \multicolumn{2}{c|}{$D_2$} \\ \hline
		
	\multicolumn{2}{|p{1.0in}<{\centering}|}{Field components} &  $a_1$ & $b_1$ & $a_2$ & $b_2$ & $a_3$ & $b_3$ & $a_4$ & $b_4$ \\ \hline
		
	\multirow{2}{0.2in}{$B_1$} &$M_1^{(1)}$ & $1$ & $1$ & $e^{i\frac{\phi_m+\pi}{4}}$ & $ie^{i\frac{3\phi_m+\pi}{4}}$ & $ie^{i\frac{\phi_m}{2}}$ & $ie^{i\frac{\phi_m}{2}}$ & $ie^{i\frac{3\phi_m+\pi}{4}}$ & $e^{i\frac{\phi_m+\pi}{4}}$ \\ \cline{2-10}
		
	&$M_2^{(1)}$ & $1$ & $-1 $& $e^{i\frac{\phi_m-\pi}{4}}$ & $e^{i\frac{3\phi_m+\pi}{4}}$ & $-ie^{i\frac{\phi_m}{2}}$ & $ie^{i\frac{\phi_m}{2}}$ & $-e^{i\frac{3\phi_m+\pi}{4}}$ & $-e^{i\frac{\phi_m-\pi}{4}}$ \\ \hline
		
	\multirow{2}{0.2in}{$B_2$} &$M_1^{(2)}$ & $1$& $-i$ & $\sqrt{2} e^{i\frac{\phi_m}{4}}$ & $i \sqrt{2} e^{i\frac{3\phi_m}{4}}$ & $e^{i\frac{\phi_m}{2}}$ & $i e^{i\frac{\phi_m}{2}}$ &\cellcolor{zcolor}  $0$ & \cellcolor{zcolor}$0$ \\ \hhline{|~|---------|}
		
	&$M_2^{(2)}$ & $1$ & $i$ &\cellcolor{zcolor}$0$& \cellcolor{zcolor}  $0$ & $-e^{i\frac{\phi_m}{2}}$ & $i e^{i\frac{\phi_m}{2}}$ & $-\sqrt{2}e^{i\frac{3\phi_m}{4}}$ & $i \sqrt{2}e^{i\frac{\phi_m}{4}}$ \\ \hline
		
	\multirow{2}{0.2in}{$B_3$} &$M_1^{(3)}$ &\cellcolor{tcolor} $2$ & \cellcolor{tcolor} $0$ &$\sqrt{2} e^{i\frac{\phi_m}{4}}$ & $i\sqrt{2} e^{i\frac{3\phi_m}{4}}$ & \cellcolor{tcolor} $0$ & \cellcolor{tcolor} $2ie^{i\frac{\phi_m}{2}}$ & $-\sqrt{2} e^{i\frac{3\phi_m}{4}}$ & $i\sqrt{2} e^{i\frac{\phi_m}{4}}$ \\ \hhline{|~|---------|}
		
	&$M_2^{(3)}$ & \cellcolor{tcolor} $0$ & \cellcolor{tcolor} $2$ & $i\sqrt{2} e^{i\frac{\phi_m}{4}}$ & $-\sqrt{2} e^{i\frac{3\phi_m}{4}}$ & \cellcolor{tcolor} $2ie^{i\frac{\phi_m}{2}}$ &\cellcolor{tcolor}  $0$ & $i\sqrt{2} e^{i\frac{3\phi_m}{4}}$ & $\sqrt{2} e^{i\frac{\phi_m}{4}}$ \\ \hline
	\end{tabular}
	\begin{tablenotes}
    \item Note: In this table, pink cells indicate the regions with traveling waves while uncolored cells denote standing waves, and blue cells indicate the regions where the field vanishes. And $\phi_m=(2m+1)\pi$, where $m$ is an integer.
   \end{tablenotes}
\end{threeparttable}
	\label{Table1}
\end{table*}

By successive substitution of the right side of each line in Eq. (\ref{Eq-Matrix_equation}) from the expression in the next line, we find:
\begin{equation} \label{Eq-Consistency}
	[a_1,b_1]^T= S_c^4 [a_1,b_1]^T= e^{i \phi} S_b^4 [a_1,b_1]^T.
\end{equation} 
The resonant modes can then be obtained by imposing a consistency condition  requiring the operator $ \exp (i \phi) S_b^4$ to have an eigenvalue equal to unity \cite{Hecht-Optics}. In order to find the values of $\phi$ that satisfy the consistency condition and hence obtain the eigenfrequencies, we first note that the eigenvalues of $S_b$ are given by $\lambda_{1,2}=\exp(\pm i \theta)$ and the corresponding eigenvectors are $\vec{v}_{1,2}=[1,\pm1]^T$. Here,  $\theta=\arcsin(\kappa) \in [0,\pi/2]$  since $\tau$ and $\kappa$ take only positive values. The consistency condition then reduces to $\exp(i\phi) \exp(\pm i4\theta)=1$, which has two solutions given by $\phi_m^\pm=\pm 4\theta+2m\pi$, where $m$ is an integer. In order to better understand this result, we recall that in the absence of beam splitters ($\kappa=0$)  the eigenfrequencies are doubly degenerate (for each resonant frequency there are two modes, one propagating in the CW and the other in the CCW direction) and are given by  $\phi_m=2m\pi$. Introducing the beam splitters results in coupling between CW and CCW modes and thus lifts the degeneracy. As a result, each degenerate pair described by $\phi_m$ splits into two modes: blue-shifted $\phi_m^+$ and red-shifted $\phi_m^-$. Interestingly, for identical 50/50 beam splitters, i.e. when $\kappa=1/\sqrt{2}$ corresponding to $\theta=\pi/4$, the eigenmodes associated with $S_b^4$ become degenerate (this is not the case for the eigenvectors of $S_b$) and hence the eigenmodes of the resonators form degenerate pairs satisfying the resonant conditions $\phi_m^+=\phi_{m+1}^-=(2m+1)\pi \equiv\phi_m$,  as shown in Figure \ref{Fig-Resonator}B. Before we proceed, we emphasize that the above-predicted degeneracy is not a result of a particular geometric symmetry. For instance, the length of any of the curved sections in the four domains ( $D_0'$, $D_0''$, $D_1$ and $D_2$) can be increased by a multiple of the operation wavelength without affecting the degeneracy despite the fact that it will break part of the geometric symmetries of the structure. 

We now investigate the eigenmode structure associated with these newly-formed degenerate modes. In principle, these eigenmodes can be expressed in any basis of the eigenvectors of $S_b^4$. Choosing a particular basis fixes the vector $[a_1,b_1]^T$ which can be then used to obtain all other field components through Eq. (\ref{Eq-Matrix_equation}). Table \ref{Table1}  lists the field values for three different bases given by: (1)  $B_1=\{\vec{v}_{1,2}\}$, $\vec{v}_{1,2}=[1,\pm 1]^T$; (2) $B_2=\{\vec{v}_{3,4}\}$, $\vec{v}_{3,4}=[1,\mp i]^T$; and (3)  $B_3=\{\vec{v}_{5,6}\}$, $\vec{v}_{5}=[2,0]^T$, $\vec{v}_{6}=[0,2]^T$. These bases are related via the linear transformations: $\vec{v}_{3,4}=\left[(1\mp i)\vec{v}_1+(1\pm i)\vec{v}_2 \right]/2$ and $\vec{v}_{5,6}=\vec{v}_1\pm\vec{v}_2$. Expressed differently, $\vec{v}_{5,6}$ can be also written as $\vec{v}_{5}=\vec{v}_3 + \vec{v}_4$ and $\vec{v}_{6}=i(\vec{v}_3 - \vec{v}_4)$.

\begin{figure*}[!t]
	\includegraphics[width=6.5in]{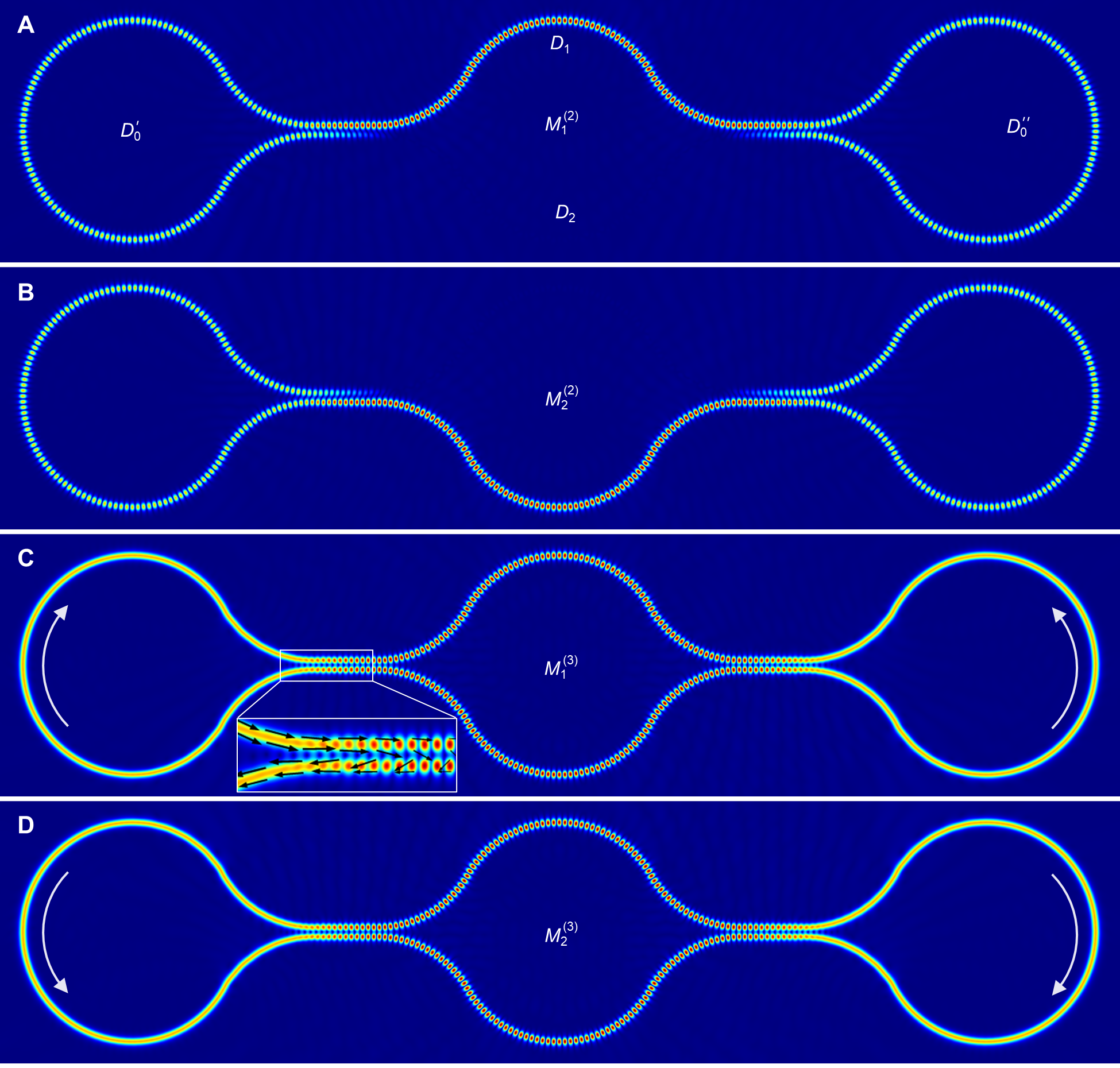}
	\caption{Eigenmodes of the hybrid-wave resonator presented in Figure \ref{Fig-Resonator}. (A) and (B) are plots of electric field component perpendicular to the resonator's plane ($|E_z|$) associated with the two degenerate standing-wave modes $M_{1,2}^{(2)}$ which resides in domain $D_0 \cup D_1$ and $D_0 \cup D_2$, respectively, where $D_0=D_0' \cup D_0''$. These corresponds to basis $B_2$. On the other hand,  (C) and (D) depict the field distribution corresponding to $M_{1}^{(3)}=M_1^{(2)} + M_2^{(2)}$ and $M_{2}^{(3)}=M_1^{(2)} - M_2^{(2)}$, which feature a hybrid standing- and traveling-wave character. The white arrows in (C) and (D) indicate the traveling direction of traveling wave. The Poynting vectors (black arrows in inset of (C)) circulate around the loop in domain $D_0$ and vanish in domain $D_{1,2}$.  The details of the geometry and material parameters used in these simulations are presented in Appendix C.}
	\label{Fig-Eigenmode}
\end{figure*}

Table \ref{Table1} lists the field components associated with the degenerate eigenmodes of the structure shown in Figure \ref{Fig-Resonator} as expressed in the three different bases $B_{1,2,3}$. The modes are expressed by the vector $M_i^{(j)} \equiv [a_1, b_1, a_2, b_2, a_3,$ $b_3, a_4, b_4]$ in each basis, where $i$ represents the mode number and $j$ denote the basis number. Note that there is a pure standing wave whenever the field components belonging to any domain have the same amplitude. On the other hand, if one of the field components vanishes, the wave is traveling. Evidently, in basis $B_3$ the eigenmodes exhibit a hybrid-wave character with both standing and traveling waves coexisting as part of the same mode. Obviously, modes $M_{1,2}^{(1)}$ represent a standing wave that extends all over the structure. On the other hand, modes $M_{1,2}^{(2)}$ represent a standing wave that covers only part of the structure. Thus modes $M_{1,2}^{(3)}$ represent potential candidates for satisfying the conditions necessary for generating hybrid-wave modes. Indeed, this is confirmed by the field distribution of modes $M_{1,2}^{(3)}$ which exhibits the dual character of traveling and standing waves covering different domains of the resonators at the same time.

An important observation here is that for the perfectly closed resonator with no loss, the hybrid-wave modes occur only when the beam splitter is $50/50$. A slight deviation from this condition removes the degeneracy and destroys the hybrid nature of the modes. This may seem to pose a challenge for experimentally observing these modes. However, realistic resonators are not perfectly closed but rather have losses due to optical absorption, radiation to free space, or due to the coupling to input and output channels. In turn, this will introduce an upper limit on the resonator's quality factor and result in a finite bandwidth of operation, which relaxes the above constraint as discussed in detail in Appendix B and C.

In order to verify the above predictions, we perform a finite element method (FEM)  full-wave simulation (using COMSOL software package) of a realistic implementation for the structure shown in Figure \ref{Fig-Eigenmode}. The details of the geometry and material parameters used in our simulations are presented in Appendix C. The eigenmodes generated by COMSOL package are in the basis $B_1$ and are shown in Appendix C. Figure \ref{Fig-Eigenmode}A and \ref{Fig-Eigenmode}B show the electric field distributions associated with the modes as represented in basis $B_2$ which are generated via linear superposition of the degenerate modes $M_1^{(1)}$ and  $M_2^{(1)}$. On the other hand, the field distributions in basis $B_3$ are depicted in Figure \ref{Fig-Eigenmode}C and \ref{Fig-Eigenmode}D. The nature of the waves can be deduced from the field distribution. Standing waves are visible through their interference pattern while traveling waves are characterized by uniform fields without interference. These plots are in agreement with the field distributions expected from Table \ref{Table1} and indeed confirm the results obtained using the scattering matrix analysis above. The field distribution of modes $M_{1,2}^{(3)}$ deserves more attention. At the center of the middle sections (domains $D_{1,2}$), the field features a standing wave while at the center of domains $D_0'$ and $D_0''$, they feature traveling waves. At the beam splitter junction, however, the field is neither a pure standing nor traveling wave. These regions represent transition domains where the wave gradually changes its character. From the Poynting vector point of view, this remarkable mode structure is enabled by the beam splitter junctions acting as interferometric mirrors for domains $D_{1,2}$ while at the same time recirculating the power incident on them from domains $D_0$ in a closed loop. Importantly, the standing-traveling wave nature observed in Figure \ref{Fig-Eigenmode}C and \ref{Fig-Eigenmode}D are characteristic of the eigenmodes of a single resonator structure and not associated with a particular steady state solution under certain engineered excitation\cite{Schmid2011PRA}. Equally important is the fact that the traveling waves in the presented structure are part of the quasi-bound state within the resonator's boundaries and not part of the leaked radiation waves outside the resonator as in the case of a finite Fabry--Perot or photonic crystal geometry for example. From a practical point of view, mapping these field distributions experimentally can be done only by using near-field probes which is possible but not an easy task. In Appendix E, we discuss a more practical scheme for accessing these modes by using input/output waveguides ports evanescently coupled to the various sections of the resonator.	

\section{Local perturbations and sensing applications} 
\begin{figure*}[!t]
	\includegraphics[width=5.5in]{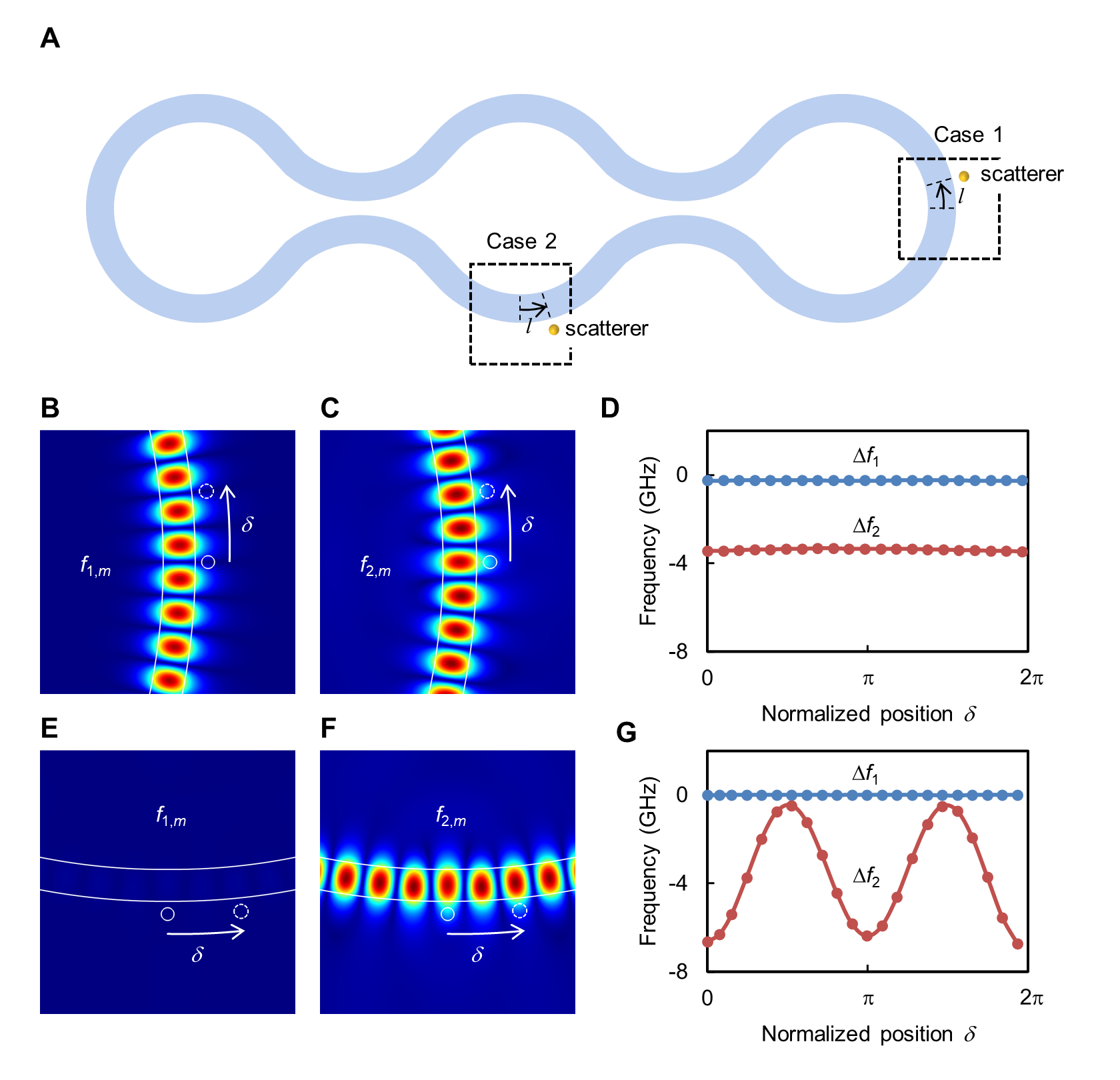}
	\caption{Effect of local perturbation. (A) A schematic of the resonator structure with nanoparticles scatterers added in the traveling wave (case 1) or standing waves (case 2) regions. In case 1, the presence of the scatterer generates two new optical modes that exhibit either a node (B) or an antinode (C) at the location of the particle with corresponding frequency shifts as shown in (D). In this case, the frequency shifts are independent of the particle location as long as it resides in the traveling wave domain. In case 2, the scatter leaves one of the modes intact with zero frequency shift (E) while at the same time introduces a perturbation to the second mode (F) with a frequency shift that varies with the location of the particle as expected (G). In both cases, $\delta=2\pi$ corresponds to a distance of 0.5 $\mu$m along the perimeter. A rigorous derivation of these results as well as their intuitive explanations are discussed in the main text. In the figure, $\Delta f_j=f_{j,m}-f_m$ with $j=1,2$. The scatterers in panels (B), (C), (E) and (F) are indicated by small white circles. }
	\label{Fig-Particle-frequency}
\end{figure*}

In this section, we investigate the effect of local perturbation due to a small scatterer on the eigenmodes of the proposed hybrid-wave resonator---a problem relevant to sensing applications \cite{Foreman2015AOP}. Given that the modes of the resonator shown in Figure \ref{Fig-Resonator}A can be written in various bases, only one of which demonstrates the hybrid-wave character, one may wonder if this feature will have any consequences under more general conditions where the particular mode is not selectively excited. This section demonstrates that this is indeed the case. To illustrate this, we consider the situation where a scatterer (nanoparticle or a fiber tip for instance) is located within the evanescent field of the hybrid-wave resonator. In particular, we investigate the two scenarios where the scatterer is located either in the traveling- or standing-wave domains, as shown in Figure \ref{Fig-Particle-frequency}A.

Before we proceed, it is useful to review the situation for purely traveling-wave resonators (such as microring and microdisk arrangements) and purely standing-wave resonators (such as Bragg and photonic crystal arrangements). In the former, the scatterer breaks the rotational symmetry of the geometry and introduces coupling between the clockwise and counterclockwise modes, leading to a splitting of the eigenfrequency \cite{Zhu2017PR,Zhu2010NP,Ozdemir2014PNAS}. Importantly, this behavior is independent of the location of the scatterer. In the latter case, however, the situation is quite different. An optical mode that has an electric field node at the scatterer location along the resonator direction will not be affected by its presence. On the other hand, a mode that exhibits an antinode at the location of the scatterer will experience a shift in its eigenfrequency \cite{Lalouat2007PRB,He2013NJP}. 

To this end, we consider a small perturbation caused by a scatterer having a scattering matrix (Appendix F): 
\begin{equation} \label{Eq-Sp}
	S_p=e^{i\phi_p}\begin{bmatrix}
		t & i r\\
		i r & t
	\end{bmatrix},
\end{equation}
where $r$ and $t$ are reflection and transmission coefficients which are taken to be real numbers satisfying $r^2+t^2=1$ (i.e. no loss); and $\phi_p=\arcsin(r)$ is  an overall additional phase.

\textbf{Scatterer located along the traveling-wave domain:}
Here, we assume a scatterer located along the traveling-wave domain, say $D''_0$, at a fixed distance from the resonator waveguide edge, as shown case 1 in Figure \ref{Fig-Particle-frequency}A. The angular position of the scatterer is defined by distance $l$ (Figure \ref{Fig-Particle-frequency}A) and the corresponding phase shift $\delta\equiv 2\pi n_\text{eff} l f/c$. By following the same approach used in deriving Eq. (\ref{Eq-Matrix_equation}), we find that the resonant frequencies (see Appendix G):
\begin{equation}\label{Eq-Particle-traveling}
	\begin{cases}
		\phi_{1,m}=\phi_m, \\
		\phi_{2,m}=\phi_m-2\phi_p.\\
	\end{cases}
\end{equation}
where $\phi_{j,m}$ with $j=1,2$ indicates the new eigenfrequencies (due to the perturbation introduced by the scatterer) branched from the unperturbed eigenfrequency $\phi_m$. In other words, the perturbation shifts the frequency of only one mode while leaving that of the other unchanged. This can be explained by the fact that the new modes, arising because of the perturbation, both exhibit a standing-wave pattern, with the node of one mode and the antinode of the other located at the position of the scatterer. Moreover, Eq. (\ref{Eq-Particle-traveling}) does not depend on the  location of the scatterer along the perimeter of the resonator, as long as it lies in the traveling-wave domain. This behavior is exactly identical to the case of a scatterer introduced in the vicinity of traveling-wave resonator such as a microring or microdisk geometry \cite{He2013NJP,Zhu2010OE}.  

These predictions are confirmed by performing full-wave simulations, where the scattering was introduced via a nanoparticle. Figure \ref{Fig-Particle-frequency}B and \ref{Fig-Particle-frequency}C depict the field distribution of the perturbed modes around the particle. Note that, as expected, the particle modifies the field distribution and creates a standing-wave pattern. Moreover, the node of the first mode and the antinode of the second mode coincide with the particle location along the perimeter of the resonator, which is consistent with our theoretical predictions. As a result, the eigenfrequency of the first mode remains unchanged while that of the second mode experiences a constant shift that does not depend on the particle location, as shown in Figure \ref{Fig-Particle-frequency}D. As a side note, we remark that the blue dots representing the simulation data in Figure \ref{Fig-Particle-frequency}D do not exactly coincide with the zero axis as predicted by our scattering matrix analysis but rather exhibit a small shift. This can be explained by recalling that, in our analysis, we treat the particle as a Rayleigh scatterer, whereas in reality higher-order multipole terms must be considered in order to obtain more accurate results. To confirm this, we have performed additional numerical simulations for different particle sizes and indeed observed that this frequency shift decreases as the particle size is reduced (in fact we could not resolve the frequency shift for particles with radii less than 30 nm).

\textbf{Scatterer located along the standing-wave domain:} Next, we consider the case when the scatterer is located in the standing-wave region, for instance in domain $D_2$, as shown by case 2 in Figure \ref{Fig-Particle-frequency}A. The resonant frequencies are given by (see Appendix G):
\begin{equation}\label{Eq-Particle-standing}
	\begin{cases}
		\phi_{1,m}=\phi_m, \\
		\phi_{2,m}=\phi_m-2\phi_p [1+(-1)^{m+1}\cos2\delta].\\
	\end{cases}
\end{equation}
The mode corresponding to the eigenfrequency $\phi_{1,m}$ (Figure \ref{Fig-Particle-frequency}E) is associated with $M_1^{(2)}$ in Figure \ref{Fig-Eigenmode}A, in which the electric field is zero at domain $D_2$, leading to an unperturbed resonant frequency after adding the particle at domain $D_2$. On the other hand, the eigenfrequency  $\phi_{2,m}$ corresponds to a perturbation of mode $M_2^{(2)}$ in Figure \ref{Fig-Eigenmode}A. Since  mode $M_2^{(2)}$ is a pure standing wave at domain $D_2$, the eigenfrequency $\phi_{2,m}$ varies with the angular position of the scatterer  (Figure \ref{Fig-Particle-frequency}F and \ref{Fig-Particle-frequency}G). In Eq. (\ref{Eq-Particle-standing}), when $m$ is an odd number, it is an antinode at the middle of $D_2$ domain ($\delta=0$), $\phi_{2,m}$ experiences the maximum frequency shift $-4\phi_p$ from $\phi_m$; when $m$ is an even number, it is a node at the middle of $D_2$ domain, the scatterer will not alter the field much and the resonant frequency will stay the same, i.e., $\phi_{2,m}=\phi_m$  at $\delta=0$. In both scenarios, $\phi_{2,m}$ will oscillate between $\phi_m$ an $\phi_m-4\phi_p$ as a function of $\delta$. In our simulation, the fact that $m$ is an odd number can be determined from Figure \ref{Fig-Eigenmode}, and it is also verified by the electric field around the particle are shown in  Figure \ref{Fig-Particle-frequency}F. The two eigenfrequencies varying with $\delta$ are shown in Figure \ref{Fig-Particle-frequency}G, consistent with Eq. (\ref{Eq-Particle-standing}). 

From the above analysis, it is clear that a resonator exhibiting hybrid-wave modes will respond very differently to perturbations affecting the standing- or traveling-wave zones. In terms of applications, this can be useful in a number of ways. For instance, the larger splitting in the location of the field maxima in the standing-wave zone can be utilized for selective sensing by functionalizing \cite{Vollmer2021NatRev} this exact location with receptors that can bind only to a particular molecule while at the same time use the traveling-wave zone for excitation and collection. On the other hand, one can instead use both zones for detecting the presence of more than one molecule (but only one at a time). This can be achieved by attaching different receptors to each zone and inferring the presence of a particular substance by measuring the degree of splitting. We plan to investigate these possibility in future works.

\section{Conclusion}
In conclusion, we have proposed a new concept for optical resonators that exhibit simultaneously co-existing standing and traveling waves as part of the field distribution of the same optical mode but occupying different locations along the resonator geometry. In addition, we have presented a specific example of a structure that implements this concept and verified its standing and traveling wave nature by using scattering matrix analysis and FEM full-wave simulations. We have investigated the robustness of the hybrid-wave feature and shown that the openness of the system allows for a larger bandwidth of operation and thus facilitates experimental observation. In addition, we have described a practical experiential scheme for probing the hybrid-wave nature by using several waveguide channels attached to various sections of the resonator geometry. Furthermore, we have discussed the implication of the hybrid-wave nature for sensing applications by investigating how the eigenmodes of such hybrid-wave resonator interact with a small scatterer located at different sections of the structure, demonstrating that the system's response can be very different depending on the location of the scatterer along the standing- or traveling-wave sections.  Another arena where hybrid-wave modes may prove useful is optical manipulation and trapping of particles. For instance, it is expected that a nanoparticle located in the traveling-wave zone will experience radiation pressure and lateral force acting in the direction towards the resonator, while a similar particle located in the standing-wave zone will in addition be subject to a trapping force along the perimeter of the resonator. Furthermore, actively tuning the beam-splitting values may allow for controlling the behavior of the resonator in real time and thus controlling its interaction with nanoparticles.	

In addition, the existence of the hybrid-wave modes identified above, which to the best of our knowledge has not been known before, raises several fundamental questions in photonics, nonlinear and quantum optics applications. For instance, it is not a priori clear how such a resonator will behave under nonlinear conditions. Does it exhibit different nonlinear bistability responses than those observed in conventional ring resonators \cite{Braginsky1989PLA}? Does it provide any new features in terms of frequency comb generation? Can soliton crystals \cite{Cole2017NPhot} form in the presence of hybrid-wave modes? Along similar lines, it is not clear to what extent the presence of the hybrid-wave modes will impact the dynamics and instability features of laser devices made of such resonators. In the quantum domain, it would be interesting to explore how quantum emitters located inside or in the vicinity of such resonators will behave. How would spontaneous emission and superradiance scale in different sections of the resonators? It has been shown previously that the interaction between atoms and electromagnetic waves featuring a standing field pattern depends on the type of resonator \cite{Shore1991JOSAB} (standing or traveling wave resonator). What makes these exploratory questions particularly interesting is that the proposed resonator exhibit transition regions (the beam splitter regions in Figure \ref{Fig-Resonator}A that interpolates between the traveling- and standing-wave domains). Light-matter interaction in this region is expected to differ from its typical behavior in standard traveling- and standing-wave resonators, which may lead to interesting new effects. At the engineering level, our work also raises interesting questions. For instance, is there a fundamental size limit on building hybrid-wave resonators? Can one implement a small volume hybrid-wave mode? What would be the modes of these structures when implemented in material platforms that support plasmonic resonances? We plan to investigate these open questions as well as implementations in other platforms such as acoustics \cite{Ingard1953JASA} and microwave \cite{Pozar1990Microwave} in future works.

\begin{acknowledgements}
This project is supported by the Air Force Office of Scientific Research (AFOSR) Multidisciplinary University Research Initiative	(MURI) Award on Programmable systems with non-Hermitian quantum dynamics	(Award No. FA9550-21-1-0202). R.E. also acknowledges support from Army Research Office (ARO) (Grant No. W911NF-17-1-0481), National Science Foundation (NSF) (Grant No. ECCS 1807552), Henes Center for Quantum Phenomena, and the Alexander-von-Humboldt Foundation. S.K.O. also acknowledges support from ARO (Grant No. W911NF-16-1-0013, W911NF-17-1-0481), NSF (Grant No. ECCS 1454531, DMR-1420620, ECCS 1807552, ECCS 1757025, ECCS 1807485).
\end{acknowledgements}

\appendix

\renewcommand{\thefigure}{S\arabic{figure}}

\setcounter{figure}{0}

\section{Implementation of hybrid-wave resonator using free-space optics}

\begin{figure}[t]
	\includegraphics[width=3.4in]{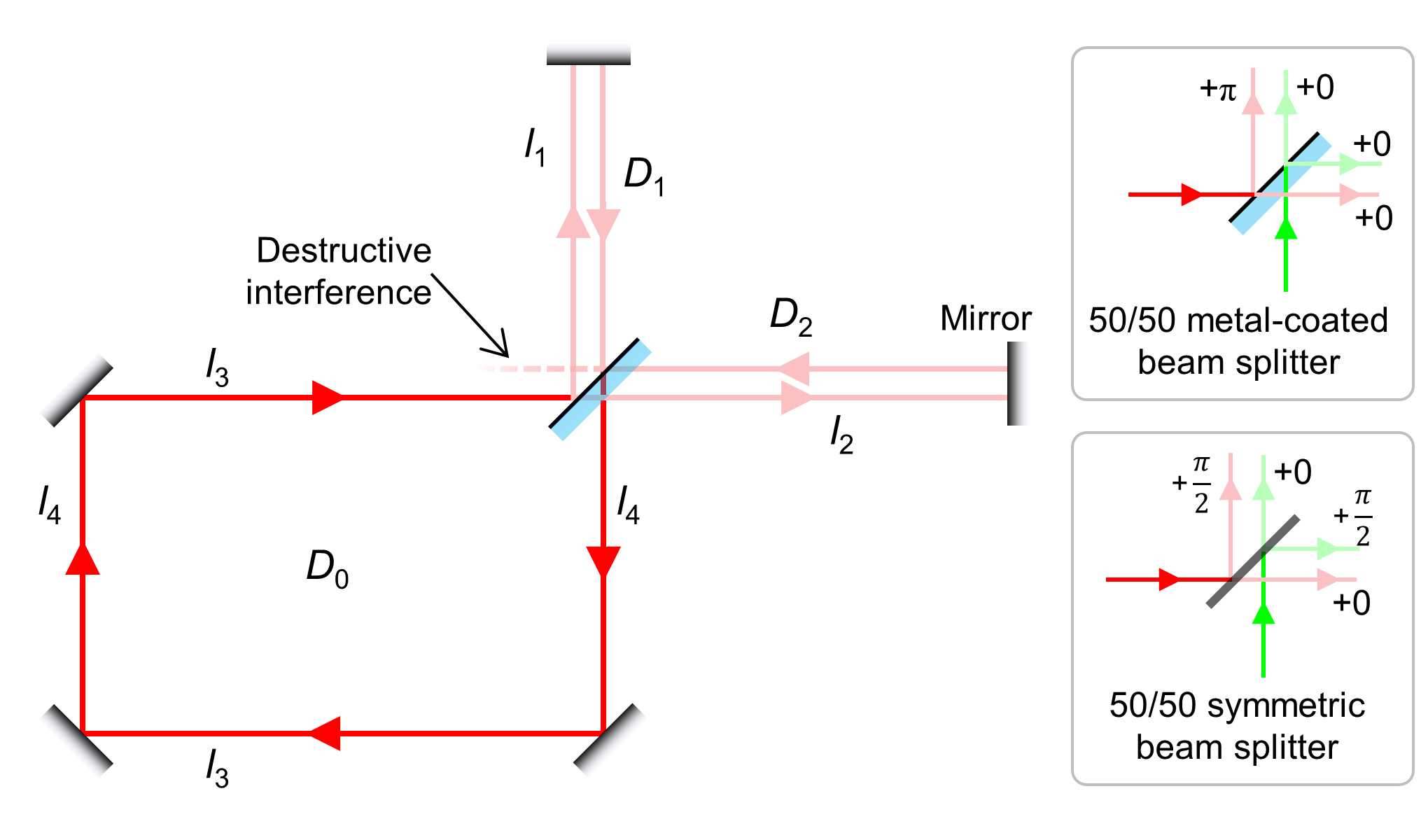}
	\caption{A setup based on free-space optics that can support a hybrid-wave mode. As discussed in the text, the parameters of the resonators can be adjusted to produce destructive/constructive interference at the back output signals from the beam splitter in order ensure the formation of a standing-wave pattern in domains $D_{1,2}$ and a traveling-wave pattern in $D_0$. Two different kinds of beam splitters, i.e., metal-coated and symmetric beam splitters, are illustrated on the right.}
	\label{Fig-Free-space}
\end{figure}

In the main text, we have discussed a possible implementation of hybrid-wave resonators using a integrated photonics platform. However, the concept of hybrid-wave resonators introduced in this work is general and can be implemented using other platforms. In this section, we demonstrate a possible realization based on free-space optics using the resonator geometry. As shown in Fig. \ref{Fig-Free-space}, it consists of five mirrors and one 50/50 beam splitter. Depending on the actual physical design of the beam splitter, it could impart different phases on the transmitted and reflected waves as explained in Fig. \ref{Fig-Free-space}. The method used in the main text can be also employed here to find the optical modes of the resonator in Fig. \ref{Fig-Free-space} for light with electric field polarized in a direction perpendicular to the plane of the resonator. Doing so, reveals the existence of two modes: the first of which, labeled as $M_1$, extends over domains $D_0 \cup D_1$ and has a resonant wavelength satisfying $2(l_1+l_3+l_4) \cdot  \frac{2\pi}{\lambda}+5\pi=2 m_1 \pi$, where $m_1$ is an integer and the additional $5\pi$ term on the left side is due to the phase shift of the beam splitter (corresponding to the metal-coated beam splitter in Fig. \ref{Fig-Free-space}) and four mirrors. The second mode, which we will refer to as $M_2$ occupies $D_0 \cup D_2$ and its resonant wavelength is given by $2(l_2+l_3+l_4) \cdot  \frac{2\pi}{\lambda}+4\pi=2 m_2 \pi$, where $m_2$ is integer and the additional $4\pi$ term is due to the phase shift of mirrors. By carefully adjusting the value of arm lengths $l_1$ and $l_2$, the resonance frequencies of these two modes can be tuned to form a degenerate pair. Following a similar argument to the discussion in the discussion in the main text, we thus see that a proper linear combination of mode $M_1$ and $M_2$ can form a traveling wave in domain $D_0$ and standing waves in domain $D_1$ and $D_2$. The light trajectory describing such a hybrid-wave mode is depicted in Fig. \ref{Fig-Free-space} by the red lines.  Intuitively, the two oppositely traveling waves in $D_{1,2}$ form a standing-wave pattern while the constructive/destructive interference at the beam splitter back output (corresponding to the condition $2(l_2-l_1)\cdot \frac{2\pi}{\lambda}=[2(m_2-m_1)+1]\pi$ ) ensures the existence of only a traveling wave in the loop section of $D_0$ is satisfied. A similar conclusion also applies to the case of  beams splitters with symmetric phase shifts such as that shown in Fig. \ref{Fig-Free-space} (for instance cube beam splitters or Pellicle beam splitters). Importantly, we note that the above discussion does not take into account free-space diffraction. Thus, it should be treated as a first-order approximation. More careful designs should be considered for actual free-space implementations.

\section{Device bandwidth}

\renewcommand{\theequation}{B\arabic{equation}}
\setcounter{equation}{0}

As discussed in the main text, in the absence of any loss mechanism, the resonator structure shown in Fig. 2A will exhibit hybrid-wave modes only for 50/50 beam splitters. A carefully designed beam splitter may indeed exhibit a 50/50 splitting ratio at a chosen operating frequency, which we call $\phi_b$ (remember that $\phi \propto f$). This however will occur over a small bandwidth. In general, the resonant frequency of the resonator may not coincide with $\phi_b$. This problem can be mitigated by tuning the length of the resonator throughout the design process to shift one of its resonant frequencies to $\phi_b$. However, any deviation in the design parameters from their ideal target values will eventually break the degeneracy of the modes. This in turn puts stringent constraints on any realistic design and can hinder any efforts for experimental observation of such an effect. Fortunately, however, realistic optical resonators are not closed systems but rather experience loss due to radiation, material absorption and coupling to input/output channels. As we will show, this will relax the above constraint.

\begin{figure}[!b]
	\includegraphics[width=3.42in]{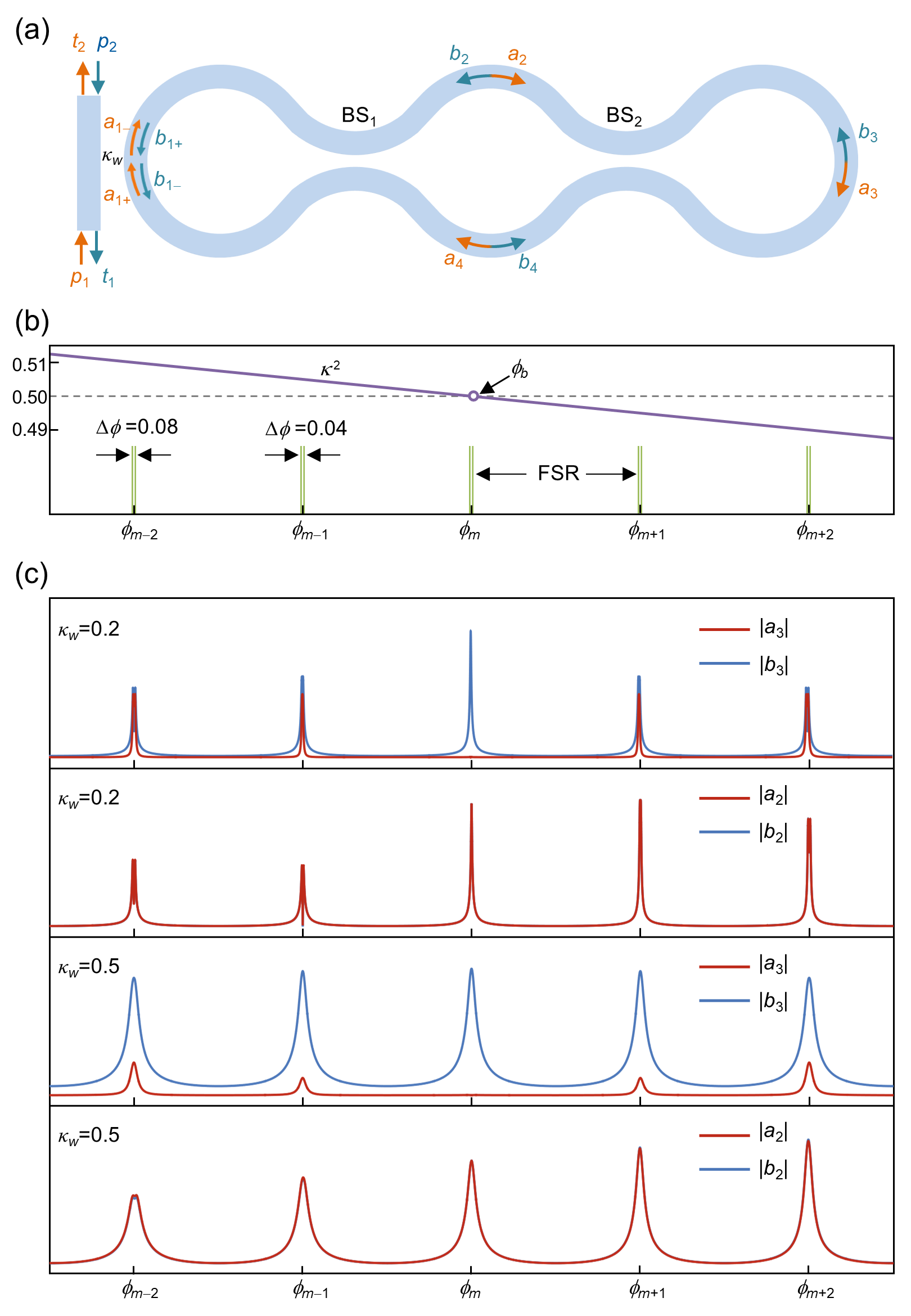}
	\caption{The bandwidth of the device. (a) A schematic of a resonator structure similar to that of Fig. 2A with an additional evanescent coupled waveguide acting as an excitation and loss channel at the same time. (b) The dependency of the beam splitter coupling coefficient $\kappa^2$ as a function of frequency (obtained with FEM simulations) shows deviation from the ideal 50/50 operating point. Here $\phi_m$ coincides with $\phi_b$ and it is the ideal 50/50 operating point. However, the degeneracy of the modes centered around $\phi_{m\pm1}$ will be lifted with $\Delta\phi=0.04$ due to the deviation of $\kappa^2$ from its ideal value of 0.5. (c) Plots of field amplitudes  $|a_{2,3}|$ and $|b_{2,3}|$ as a function of frequency (calculated by using scattering matrix formalism) for two different values of the coupling to the waveguide, $\kappa_w=0.2$ and 0.5, when the system is externally excited with $p_1=1$. As expected for a more lossy system, the hybrid-wave feature persists over a larger frequency range.}
	\label{Fig-Bandwidth}
\end{figure}

Let us consider the geometry in Fig. \ref{Fig-Bandwidth}(a)  where the resonator is evanescently coupled to a waveguide that is used for wave excitation and output signal collection. In this structure, the main sources of loss are coupling to the waveguide and radiation loss to free space modes. This latter mechanism limits the resonator's quality factor even in the absence of a waveguide, as we discuss in detail in Appendix C.  The finite element method (FEM) full-wave simulation shows that the beam splitting ratio is 50/50 only at one single frequency and can vary from 51/49 to 49/51 over five free spectral range (FSR) periods. As a result, the modes of the resonator are not degenerate as they would be desired but rather experience a splitting, as illustrated in Fig. \ref{Fig-Bandwidth}(b). To further elucidate this behavior, we consider an input excitation from the bottom port of the waveguide and we compute the field coefficients $a_2$, $b_2$ and $a_3$, $b_3$ (see Fig. \ref{Fig-Bandwidth}(a) for a full list of the field amplitudes and their respective locations along the resonator structure) by using the scattering matrix formalism:
\begin{equation}\label{Eq-wg-resonator}
	\begin{aligned}
\relax	[a_{1-}, t_2]^T &= S_w[a_{1+}, p_1]^T, \\
		[b_{1-}, t_1]^T &= S_w[b_{1+}, p_2]^T, \\
		[a_{1+}, b_{1+}]^T &= S_c[a_4, b_2]^T, \\
		[a_4, b_2]^T &= S_c [a_3, b_3]^T, \\
		[a_3, b_3]^T &= S_c [a_2, b_4]^T, \\
		[a_2, b_4]^T &= S_c [a_{1-}, b_{1-}]^T, 
	\end{aligned}
\end{equation}  
where  $S_w=\begin{bmatrix}\tau_w & i \kappa_w\\ i \kappa_w & \tau_w\end{bmatrix}$ is the scattering matrix  between the waveguide and the resonator. Here, $\tau_w$ and $\kappa_w$ are the field transmission and coupling coefficient, and they satisfy $\tau_w^2+\kappa_w^2=1$. The input amplitudes are taken to be $p_1=1$ and $p_2=0$. In the ideal scenario, $|\frac{a_2}{b_2}|=1$ and $a_3=0$. The actual values of the various field components are plotted in Fig. \ref{Fig-Bandwidth}(c) as a function of frequency for two values of the coupling coefficient: $\kappa_w=0.2$ and $\kappa_w=0.5$. As can be observed, in the former case, the standing-traveling wave feature persists only over a very narrow band centered around $\phi_b$. In the latter case, however, the field values and ratios still represent standing and traveling waves in their corresponding resonator sections, at least to a good degree of approximation to the perfect behavior in a larger frequency band around $\phi_b$. Evidently, the openness of the device thus facilitates the experimental observation of the hybrid standing and traveling wave features.

\section{Simulation details}

\begin{figure}[b]
	\includegraphics[width=3.4in]{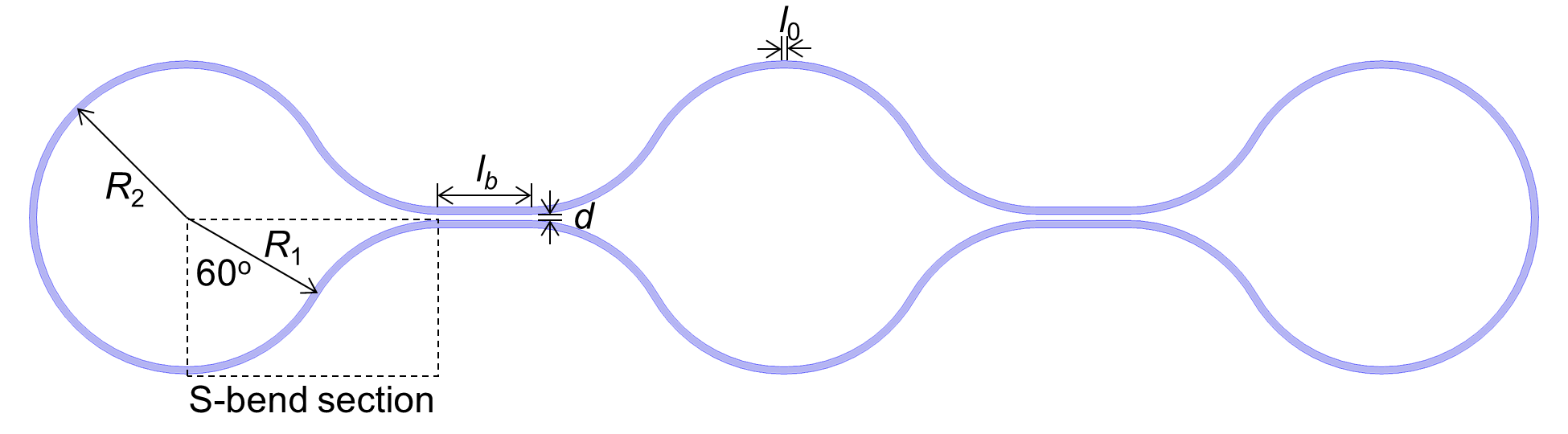}
	\caption{The geometry of the resonator structure used in our FEM simulation. The waveguide of the resonator has a width of $w=0.25$ $\mu$m, and a refractive index of 3.47, surrounded by a cladding with an index of 1.44 (typical values for silicon and silica at a telecom wavelength). The edge-to-edge gap in the directional coupler is $d=0.2\ \mu$m, and the length of the linear section in the coupling region is $l_b=3\ \mu$m. The S-bend section is constructed by connecting two 60$^\circ$ arcs with radius $R_1=5\ \mu$m. The top and lower S-bend sections on each side of the resonator are then connected by a half-circle with $R_2=R_1+w/2+d/2=5.225\ \mu$m. A linear section with length $l_0 =0.08\ \mu$m is inserted into the top and lower middle sections in order to tune the eigenfrequency of the resonator equal to the frequency $f_b$ at which the splitting ratio of the beam splitters is 50/50.}
	\label{Fig-Geometry}
\end{figure}

\begin{figure}[h]
	\includegraphics[width=3.4in]{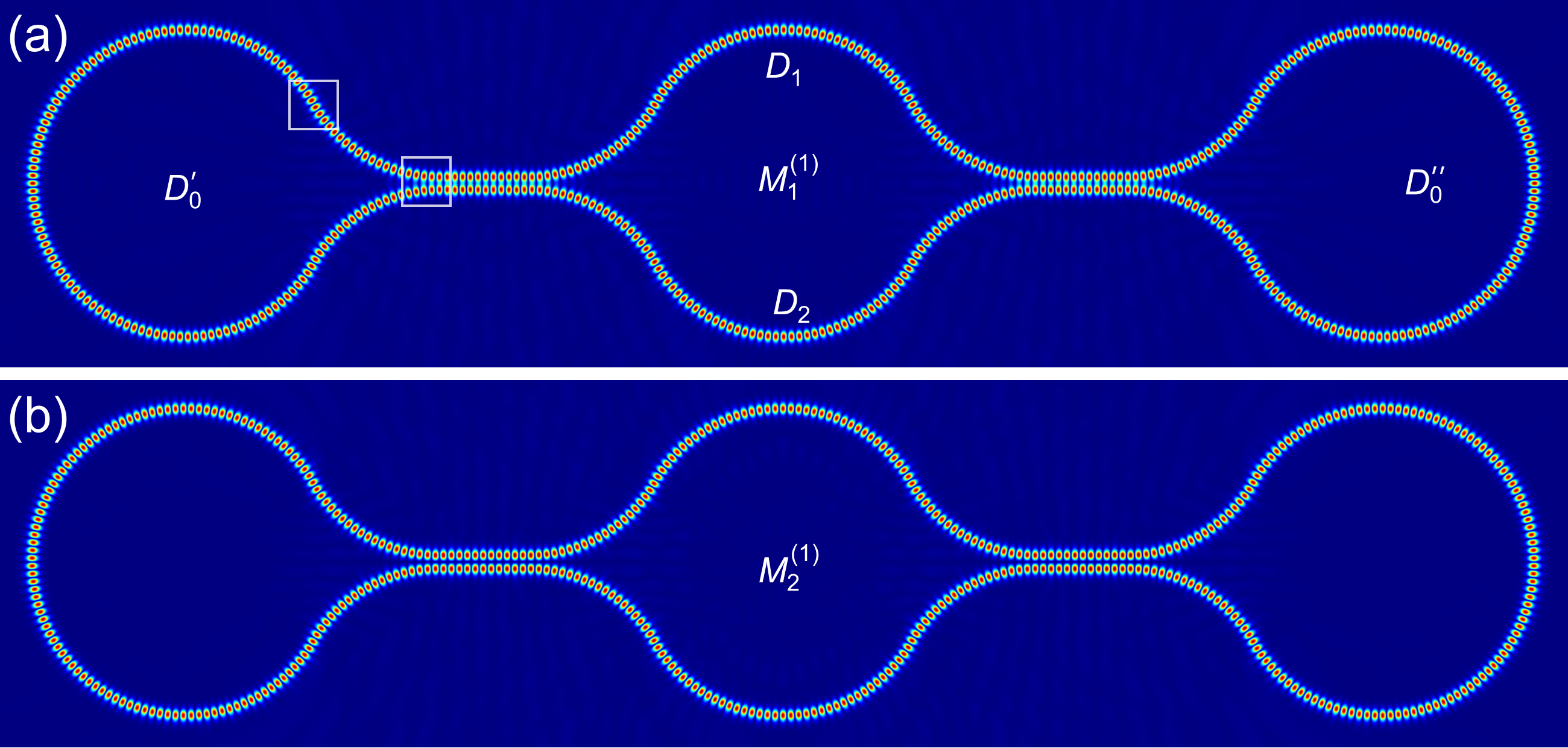}
	\caption{COMSOL-generated electric field ($|E_z|$) distributions  associated with the eigenmodes $M_{1,2}^{(1)}$ in bases $B_1$.}
	\label{Fig-Eigenmode12}
\end{figure}

The simulation of eigenmodes in Fig. 3 in the main text is performed by using frequency-domain FEM available via COMSOL package. The geometry is shown in Fig. \ref{Fig-Geometry} and the parameters are listed in the figure caption. For fixed design parameters, the frequency ($f_b$) at which the beam splitting is 50/50, can be tuned by changing the beam splitter coupling length $l_b$. For example, in our simulations we take $l_b=3\ \mu$m in order to obtain $f_b = 192.1$THz (corresponding to a wavelength of 1.56 $\mu$m). In addition, a linear section of length $l_0$ is inserted into the top and lower sections to act as a design knob for controlling the location of the eigenfrequency of the resonator, ensuring that it coincides or at least overlaps with $f_b$. In our design, we take $l_0 =0.08\ \mu$m. For these parameters, FEM simulations indeed show that the two eigenmodes are almost degenerate with frequencies $f^+=192.11649$ THz and $f^-=192.11647$ THz, as shown in Fig. \ref{Fig-Eigenmode12}. These two eigenmodes are $M_{1,2}^{(1)}$ in bases $B_1$. Due to the degenerate nature of these modes, any linear combination is also a mode. The field profiles plotted in Fig. 3 are constructed by using such appropriate superpositions as discussed in detail in the main text. 

In order to evaluate the quality factor of the isolated resonator (i.e. without coupling to waveguides), we also extract the imaginary component of the resonant frequency from the FEM simulations. For the two degenerate modes in Fig. \ref{Fig-Eigenmode12}, the value of this imaginary component is $\sim1$ GHz, which results in a quality factor of  $Q\equiv\dfrac{f'}{2f''}\approx \dfrac{192\ \text{THz}}{2 \times 1\ \text{GHz}}= 10^5$. Further simulations indicate that the main source of the radiation loss in our structure are the S-bend sections and their connections to the beam splitter domains. For clarity, we illustrate two of these sections in Fig. \ref{Fig-Eigenmode12}(a). This is, however, not a fundamental source of loss and it is possible that further geometric optimization can reduce this loss and boost the quality factor.

\section{An alternative design using integrated photonic platforms} 
\begin{figure} [!t]
	\includegraphics[width=3.4in]{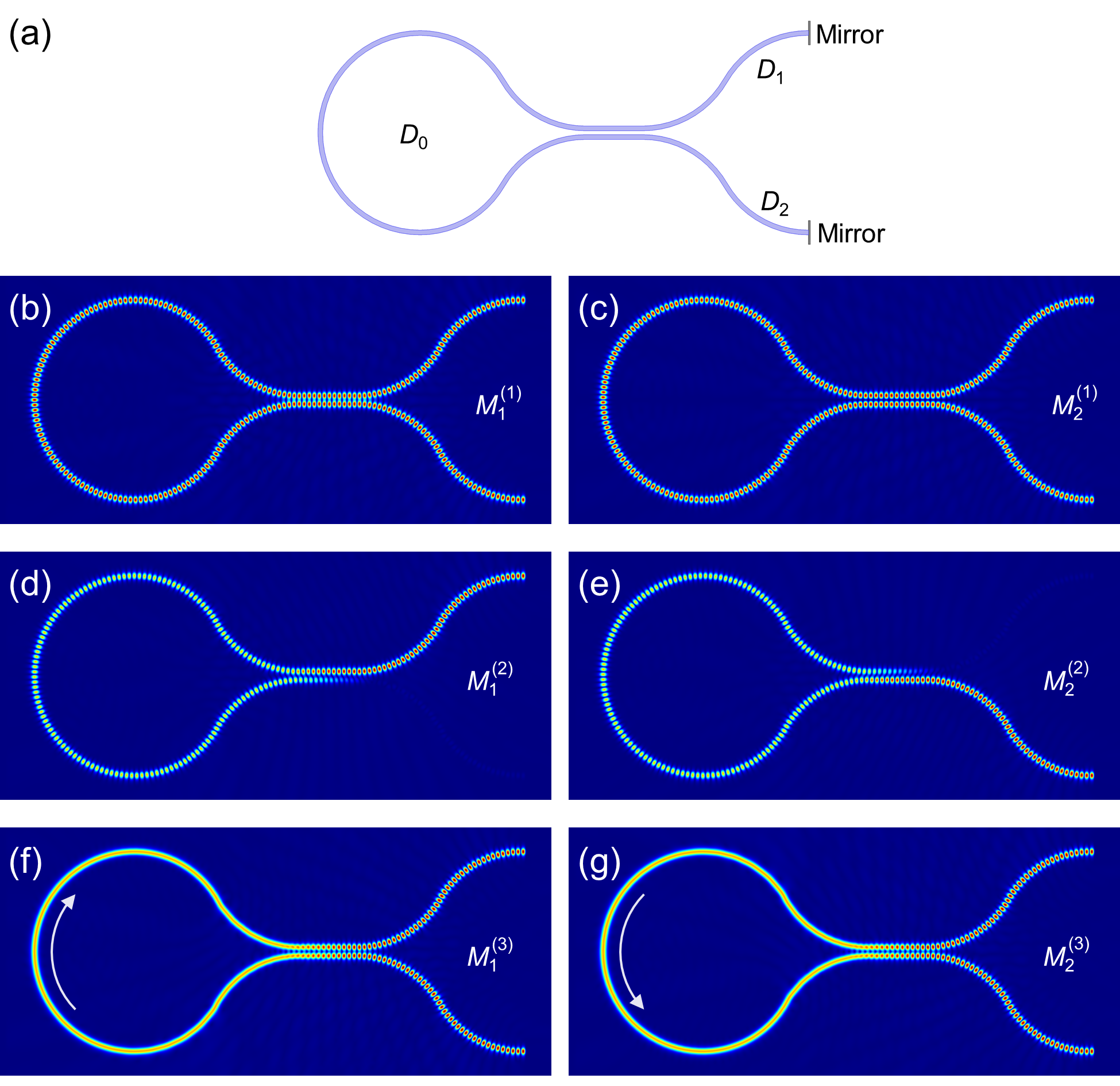}
	\caption{A hybrid-wave resonator construed with mirrors. (a) The structure similar to that proposed in the main text after replacing the right most loop mirror with two conventional mirrors.  The electric field distribution associated with the two degenerate eigenmodes (numerically obtained eigenfrequencies are $f^+=192.12172\ \text{THz}$ and $f^-=192.12165\ \text{THz}$) is plotted in different bases (b)--(g). The hybrid-wave nature is observed in the bases $B_3$, shown in (f) and (g).}
	\label{Fig-Resonator-mirror}
\end{figure}

The implementation presented in the main text is based on two loop mirrors. However, we note that this is not necessary. For instance, one can replace one of these loop mirrors with two ordinary mirrors as shown in Fig. \ref{Fig-Resonator-mirror}(a). By adopting the same design parameters as those used in Fig. \ref{Fig-Geometry}  and using full-wave simulations, we can compute the two degenerate modes of the systems. Figures \ref{Fig-Resonator-mirror} (b)--(g) present plots of these modes in different bases, where we clearly observe the hybrid traveling-standing wave nature in Fig. \ref{Fig-Resonator-mirror} (f) and (g). Here the mirror was introduced by using a 100 nm layer of silver.

\section{Experimental implementation of observing hybrid-wave modes}

\begin{figure*}[t]
	\includegraphics[width=5in]{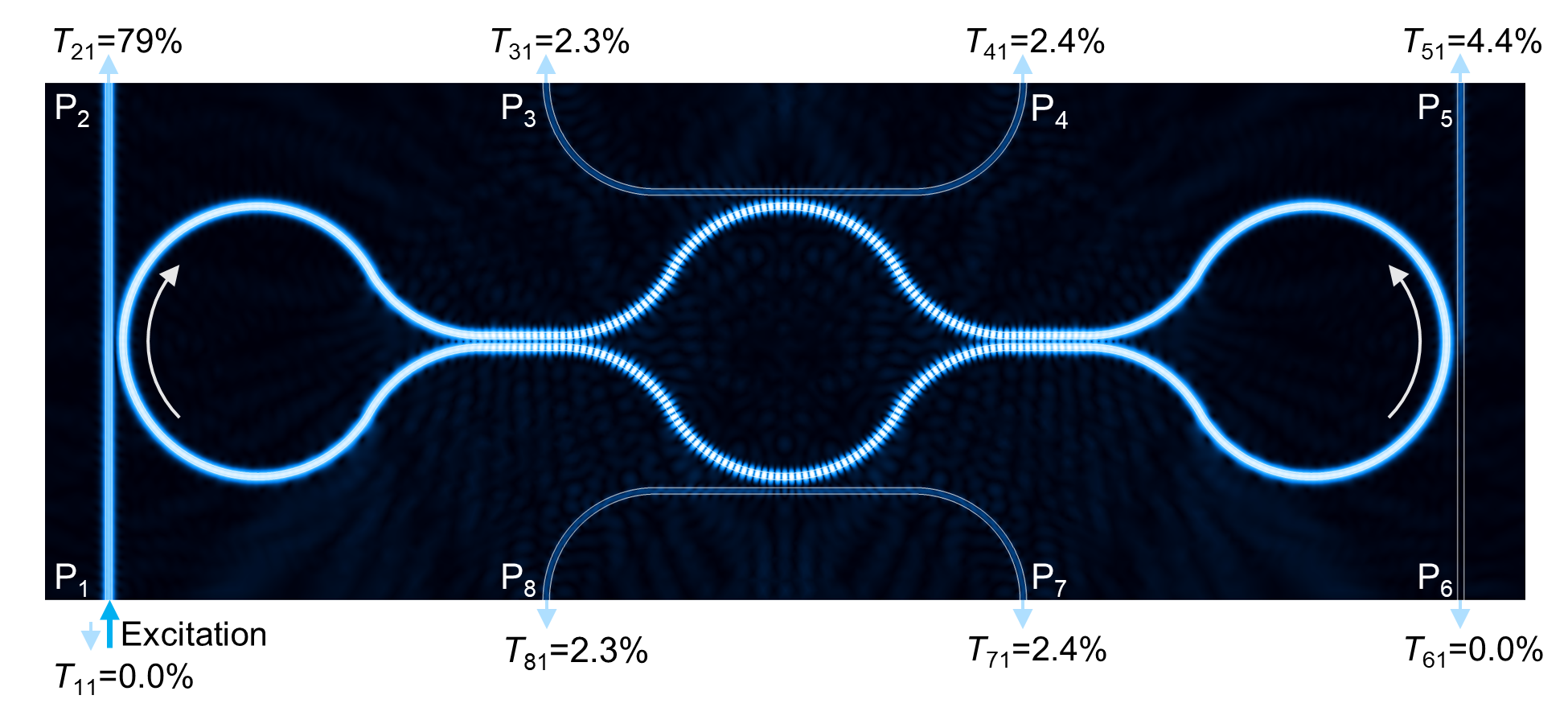}
	\caption{ The transmission of the resonator. A resonator structure similar to that of Fig. 3  with additional waveguide channels evanescently coupled at different sections. The standing or traveling wave nature of the modes can be probed by launching light into one of the waveguide ports (port $P_1$ in this particular case) and monitoring the output light from other ports. The values of the various power transmission coefficients are also listed on the figure. For example, the fact that $T_{31} \approx T_{41}$ indicates that the wave forms a standing-wave pattern in the central top section while $T_{61}=0$ indicates that the wave component at the right side of the resonator is traveling. Similar arguments can be made about other domains. Color scale represents electric field component perpendicular to the resonator's plane ($|E_z|$).}
	\label{Fig-Transmission}
\end{figure*}

In this section, we discuss how the hybrid-wave feature of the optical modes associated with the resonator geometry presented in the main text can be observed experimentally. First, owing to various possible representations of the modes due to degeneracy, one must ensure that the eigenmodes are excited in the correct bases, $B_3$ in our case. Second, one must verify the wave nature (standing or traveling) in each domain. In principle, this latter step can be performed by directly measuring the electric field distribution. However, this requires near-field measurements \cite{Jackson1995U,Abashin2006OE,Ziegler2017Np},  which is not a trivial task. A modification of the above structure, however, can provide a means for probing the modal nature in a straightforward fashion by coupling each section to an external waveguide channel as shown in Fig. \ref{Fig-Transmission}. The input and output ports of the waveguides are denoted by $P_n$ with $n=1,2,...,8$. The power transmission coefficients from port $P_m$ to $P_n$ are labeled as $T_{nm}$.

Evidently, the eigenmode $M_1^{(3)}$ can be launched by exciting the resonator from ports $P_1$ or $P_6$ and setting all other inputs to zero. Similarly, mode $M_2^{(3)}$ can be excited via the ports $P_2$ or $P_5$. Here we focus on the eigenmode $M_1^{(3)}$ and we consider an excitation from $P_1$ only, keeping in mind that the same discussion applies to $M_2^{(3)}$. Due to the traveling-wave nature in domain $D_0$, it is expected to have $T_{11}=T_{61}=0$. On the other hand, because the field distributions in domains $D_{1,2}$ correspond to a standing wave, we expect $T_{31}=T_{41} \neq 0$ and $T_{71}=T_{81} \neq 0$. This setup can thus provide direct information on the wave nature of the eigenmode in all domains without the need for any near-field measurements. These predictions are confirmed by performing FEM full-wave simulations using this modified structure.

Figure \ref{Fig-Transmission} also lists the power transmission coefficients $T_{n1}$ at each output port due to an excitation from port $P_1$ as obtained by the FEM simulations as well as field distribution as obtained by FEM simulations where the traveling- and standing-wave patterns in domains $D_0$ and $D_{1,2}$ can be observed. Evidently, these results are consistent with the field distribution of eigenmode $M_1^{(3)}$.  Note that $T_{21}$ is much larger than $T_{51}$ because output port $P_2$ is directly fed from input port $P_1$ as well as from the resonator, while port $P_5$ is fed only from the resonator. Finally, we would like to note that this strategy can also allow us to investigate the formation of exceptional surfaces \cite{Zhong2019PRL,Soleymani2022NC} in these hybrid-wave resonators which we plan to do in future works.

\section{Scattering matrix for a scatterer} 

\renewcommand{\theequation}{F\arabic{equation}}
\setcounter{equation}{0}

\begin{figure}[!h]
	\centering
	\includegraphics[width=3in]{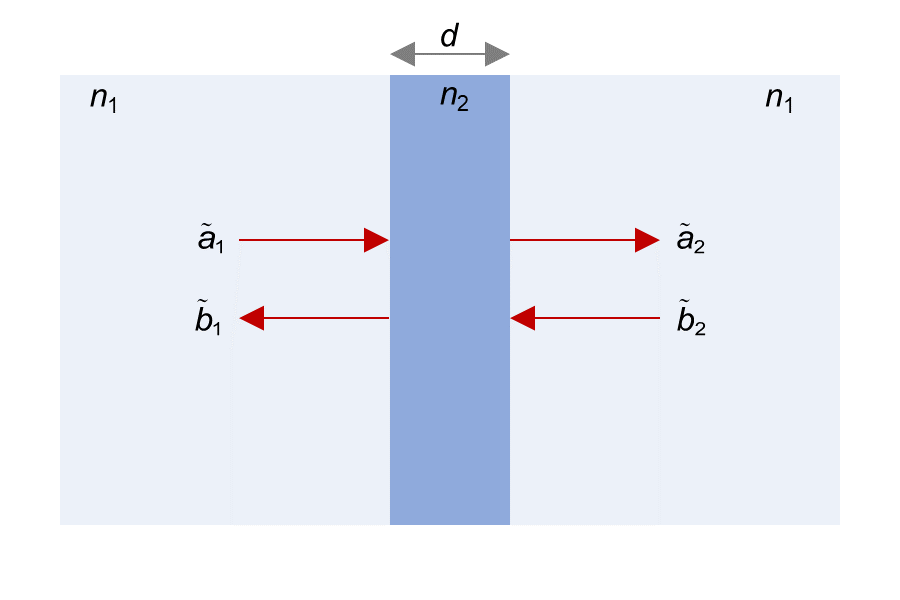}
	\caption{A schematic diagram of the scatterer geometry used in our analysis in this section. It consists of a small slab made of refractive index $n_2$ embedded in a background of index $n_1$. In the limit of $n_2/n_1 \gg1$ and $d\ll\lambda$, where $\lambda$ is the light wavelength, we can derive the expression used for the scattering matrix in Eq. (\ref{Eq:S}). As shown in the main text, this form of the scattering matrix gives excellent agreement with results obtained using FEM simulations for nanoparticle scatterers. This can be understood by recalling that in the Rayleigh regime, all small scatterers basically behave identical regardless of the details of their shapes. }
	\label{Fig-Sp}
\end{figure}

Here we derive the scattering matrix $S_p$ of Eq. (3)  which is used to describe the scattering from a small scatterer. In the main text, we considered a nanoparticle as a scatterer but in the Rayleigh regime (i.e., the size of the particle is much smaller than the wavelength of the resonant light), all small scatterers behave in the same way regardless of their shape. Thus, to facilitate the analysis, here we will consider a small slab layer made of a different refractive index than the background index as a scatterer as shown in Fig. \ref{Fig-Sp}. It is straightforward to show that the field amplitudes shown in Fig. \ref{Fig-Sp} are related by the transfer matrix equation \cite{Saleh-FP}:

\begin{equation}
\begin{split}
		\begin{bmatrix} \tilde{a}_2 \\ \tilde{b}_2 \end{bmatrix} =
		\dfrac{1}{2n_1}
		\begin{bmatrix} n_1+n_2 & n_1-n_2 \\ n_1-n_2 & n_1+n_2\end{bmatrix} 
		\begin{bmatrix} e^{i k d} & 0 \\ 0 & e^{-i k d}\end{bmatrix} \\
		\times \dfrac{1}{2n_2}
		\begin{bmatrix} n_1+n_2 & n_2-n_1 \\ n_2-n_1 & n_1+n_2\end{bmatrix} 
		\begin{bmatrix} \tilde{a}_1 \\ \tilde{b}_1 \end{bmatrix}.
\end{split}
\end{equation}
Here, $n_1$ is the background refractive index and $n_2$ is the slab index, $k=2\pi n_2/\lambda$ is the wave vector in slab and $d$ is the slab thickness.  

By substituting $\tilde{n}=n_2/n_1$ and $\theta=kd$, the total transfer matrix takes the form:
\begin{equation}
	\mathcal{T}=\dfrac{\sin\theta}{2\tilde{n}}
	\begin{bmatrix} 2\tilde{n} \cot\theta+i(\tilde{n}^2+1) & i(\tilde{n}^2-1) \\ -i(\tilde{n}^2-1)  & 2\tilde{n} \cot\theta-i(\tilde{n}^2+1) \end{bmatrix}.
\end{equation}
We now consider the following limit for the scatterer parameters:  $\tilde{n}^2 \gg 1$, $d \sim 0$ and $\theta \sim 0$. Under these conditions, $\mathcal{T}$ is given by:  
\begin{equation}
	\mathcal{T}=\begin{bmatrix} 
		1+\dfrac{i}{2}\tilde{n} \theta & \dfrac{i}{2}\tilde{n} \theta\\
		-\dfrac{i}{2}\tilde{n} \theta & 1-\dfrac{i}{2}\tilde{n} \theta
	\end{bmatrix}.
\end{equation}
Importantly, in this limit, the matrix $\mathcal{T}$ still satisfies the power conservation. By using the relation between transfer and scattering matrices, we can finally obtain an expression for the scattering matrix $S$ of the scatterer: 
\begin{equation}
	S=\dfrac{1}{1-i h}\begin{bmatrix}
		1 & ih \\
		ih & 1
	\end{bmatrix},
\end{equation}
where $h=\tilde{n}\theta/2$. Finally, by setting $r=\dfrac{h}{\sqrt{1+h^2}}$ and $t=\dfrac{1}{\sqrt{1+h^2}}$, we obtain
\begin{equation}\label{Eq:S}
	S=e^{i \phi}\begin{bmatrix}
		t & ir \\
		ir & t
	\end{bmatrix}, \text{and } \phi=\arcsin (r).
\end{equation}
This is exactly the expression used in Eq. (3) and it gives consistent results with FEM simulations. Finally, we emphasize that this simple derivation presented here can be repeated for specific nanoparticle geometries but with more involved analysis. For details, see Refs. \cite{Venugopalan1993,Bohren-ASLSP}.

\section{Proof of particle scattering} \renewcommand{\theequation}{G\arabic{equation}}
\setcounter{equation}{0}

\begin{figure*}[t]
	\includegraphics[width=4.5in]{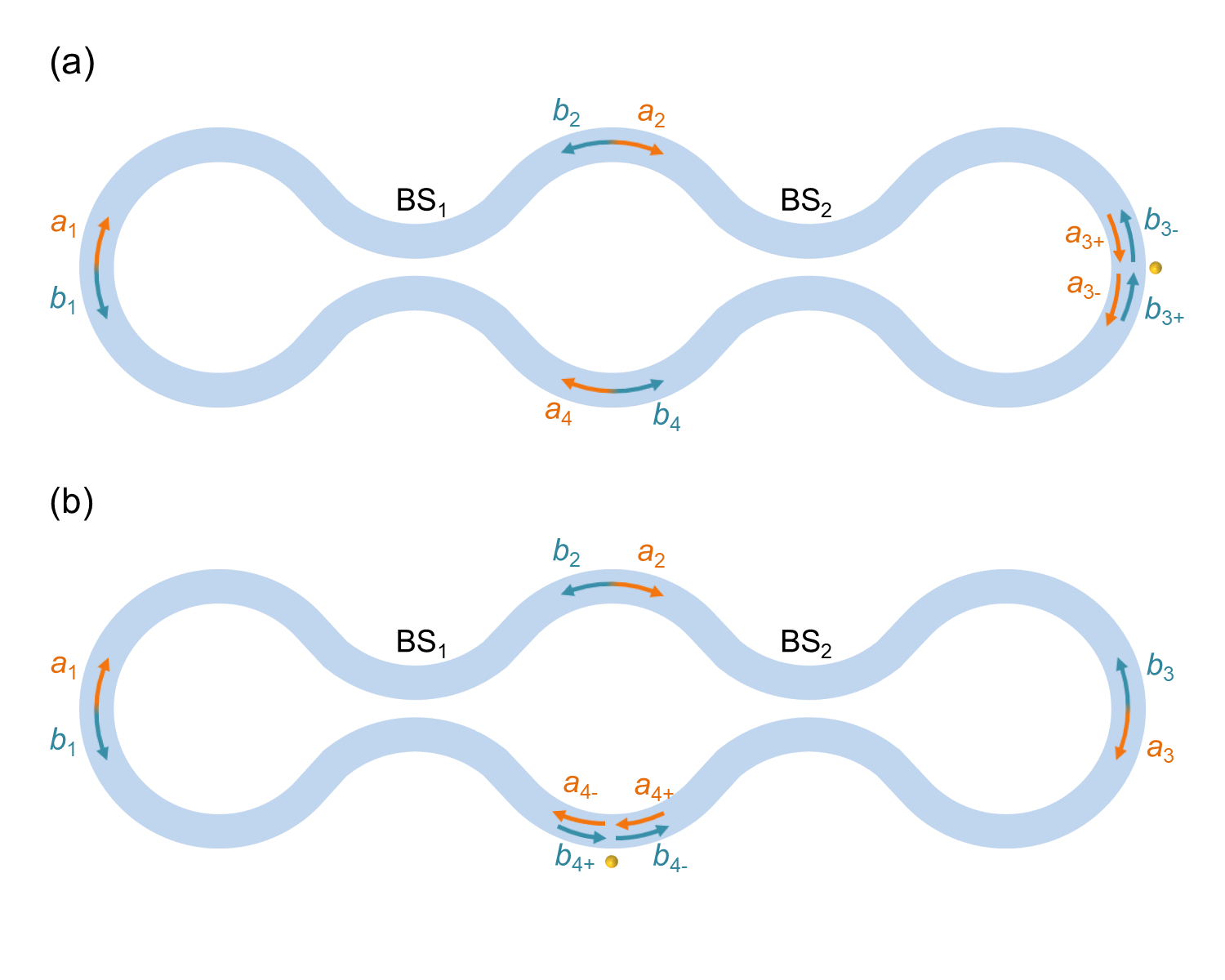}
	\caption{A schematic of the resonator structure proposed in our work showing a nanoparticle serving as a perturbation.  The particle can be located: (a) along the traveling-wave section; and (b) along the standing-wave section. Various field amplitudes used in our analysis are also depicted on the figure.}
	\label{Fig-Particle-schematic}
\end{figure*}

In this section, we present a detailed derivation for Eq. (4) and Eq. (5) which describe the eigenfrequency splitting due to a perturbation by a small scatterer located at the traveling- and standing-wave sections of the resonator, respectively (see Fig. \ref{Fig-Particle-schematic}). \\

\textbf{Scatterer located along the traveling-wave domain}: Similar to the analysis of the resonant frequency in the main text, the relation between the field amplitudes in the presence of a scatterer located in the traveling-wave section (Fig. \ref{Fig-Particle-schematic}(a)) are given by:
\begin{equation}\label{Eq-S-traveling}
	\begin{aligned}
\relax	[a_1, b_1]^T &= S_c [a_4, b_2]^T, \\
		[a_4, b_2]^T &= S_c S_{\delta}^{-1} [a_{3-}, b_{3-}]^T, \\
		[a_{3-}, b_{3-}]^T&=S_p  [a_{3+}, b_{3+}]^T, \\
		[a_{3+}, b_{3+}]^T &= S_{\delta} S_c [a_2, b_4]^T, \\
		[a_2, b_4]^T &= S_c [a_1, b_1]^T. 
	\end{aligned}
\end{equation}  
Here, $S_{\delta}=\begin{bmatrix} e^{-i \delta} & 0 \\0 & 	e^{i \delta}\end{bmatrix}$ and $\delta$ characterize  the angular position of the particle across the perimeter of te ring which is taken to be positive for the counterclockwise displacement.  After some algebraic manipulations, we obtain:
\begin{equation}
	[a_3^+,b_3^+]^T= S_{tr} [a_3^+,b_3^+]^T, 
\end{equation} 
where
\begin{equation}
	S_{tr}=S_{\delta}S_c^4 S_{\delta}^{-1} S_p =-e^{i(\phi+\phi_p)} \begin{bmatrix} t & i r  \\i r  & t\end{bmatrix}. 
\end{equation} 
The eigenvalues and eigenvectors associated with the matrix $S_{tr}$ are given by $\lambda_1=-\exp(i \phi)$, $\vec{u}_1=[1,-1]^T$ and $\lambda_2=-\exp[i (\phi+2\phi_p)]$, $\vec{u}_2=[1,1]^T$. Finally by applying the consistency condition for obtaining the eigenfrequencies of the resonator, we find the solutions for $\phi$:
\begin{equation} \label{Eq_App_phi_traveling}
	\begin{cases}
		\phi_{1,m}=\phi_m, \\
		\phi_{2,m}=\phi_m-2\phi_p.\\
	\end{cases}
\end{equation}
As expected from our discussion in the main text, only one eigenmode is affected by the perturbation. Moreover, the frequency shift does not rely on the actual angular position of the particle ($\delta$ is absent from Eq. (\ref{Eq_App_phi_traveling})). These values for the new eigenfrequencies are consistent with the field distribution of the new modes. In particular, the field associated with mode $\phi_{1,m}$ corresponding to the eigenvector $\vec{u}_1=[1,-1]^T=[a_3^+, b_3^+]^T$ has a node at the location of the particle, while that of mode $\phi_{2,m}$ with an eigenvector is $\vec{u}_2=[1,1]^T=[a_3^+, b_3^+]^T$ has an antinode at particle's position.\\

\textbf{Scatterer located along the standing-wave domain}: Next we consider the case depicted in Fig. \ref{Fig-Particle-schematic}(b) where the particle is located in the standing-wave section. In this case, the field amplitudes are related by:
\begin{equation}\label{Eq-S-standing}
	\begin{aligned}
\relax	[a_1, b_1]^T &= S_c [a_{4-} e^{i\delta}, b_2]^T, \\
		[a_{4+} e^{-i\delta}, b_2]^T &= S_c  [a_3, b_3]^T, \\
		[a_3, b_3]^T&= S_c [a_2, b_{4-}e^{-i\delta}]^T,\\
		[a_2, b_{4+}e^{-i\delta}]^T&= S_c [a_1, b_1]^T,\\
		[a_{4-}, b_{4-}]^T&=S_p  [a_{4+}, b_{4+}]^T.
	\end{aligned}
\end{equation}  
Interestingly, Eq. (\ref{Eq-S-standing}) does not lead to a simple consistency condition in the form  $[a_1,b_1]^T=S_{st}[a_1,b_1]^T$. However, by eliminating $a_{1,3}$ and $b_{1,3}$, one obtains
\begin{equation}\label{Eq-S-standing-2}
	\begin{aligned}
\relax		[a_{4+} e^{-i\delta}, b_2]^T &= S_c^2  [a_2, b_{4-}e^{-i\delta}]^T, \\
		[a_2, b_{4+}e^{-i\delta}]^T&= S_c^2 [a_{4-} e^{i\delta}, b_2]^T,\\
		[a_{4-}, b_{4-}]^T&=S_p  [a_{4+}, b_{4+}]^T.
	\end{aligned}
\end{equation}  
By noting that $S_c^2=i \exp(i \phi/2) \begin{bmatrix} 0 & 1 \\1 & 	0\end{bmatrix}$, we can finally obtain two independent consistency conditions: 
\begin{equation}\label{Eq-S-standing-3}
	\begin{aligned}
		\begin{bmatrix} a_2 \\ b_2 \end{bmatrix} &= i e^{i \frac{\phi}{2}} \begin{bmatrix} 0 & 1 \\1 & 	0\end{bmatrix} \begin{bmatrix} a_2 \\ b_2 \end{bmatrix}, \\
		\begin{bmatrix} a_{4+} \\ b_{4+} \end{bmatrix} &=i e^{i \frac{\phi}{2}} \begin{bmatrix} 0 & e^{-i2\delta}  \\e^{i2\delta} & 	0\end{bmatrix} S_p	\begin{bmatrix} a_{4+} \\ b_{4+} \end{bmatrix} .
	\end{aligned}
\end{equation}  
It is worth commenting on these last relations. These are two independent conditions that describe independent sets of modes. This feature can be understood by referring to Fig. 3A and 3B where the modes are described in bases $B_2$. In this bases, the field of mode $M_1^{(2)}$ vanishes in the lower middle section while that of $M_2^{(2)}$ vanishes in the top middle section. As a result, a nanoparticle located as shown in Fig. \ref{Fig-Particle-schematic} will affect only $M_2^{(2)}$ while at the same time leaving $M_2^{(1)}$ intact. This observation explains why two independent conditions for the eigenfrequencies arise in our analysis. By solving the above equation, we finally find the solutions for $\phi$:
\begin{equation}\label{Eq-APP-Particle-standing}
	\begin{cases}
		\phi_{1,m}=\phi_m, \\
		\phi_{2,m}=\phi_m-2\phi_p [1+(-1)^{m+1}\cos2\delta],\\
	\end{cases}
\end{equation}
which completes the proof.

\bibliography{Reference}

\end{document}